\documentclass[lettersize,journal]{IEEEtran}
\usepackage{pifont}
\usepackage{amsmath,amssymb}
\usepackage{amsthm}
\usepackage{xcolor,xspace}
\usepackage[ruled, linesnumbered]{algorithm2e}
\usepackage{tikz}   
\usepackage{url}      
\usepackage{tabularx}
\usepackage{arydshln}
\usepackage{subcaption}     
\usepackage{float}
\usepackage{multirow}
\usepackage{inconsolata}   
\usepackage{mdframed}  
\usepackage{enumitem}
\usepackage{booktabs}
\usepackage{diagbox}
\usetikzlibrary{arrows, patterns, automata, positioning}

\newmdtheoremenv[
  backgroundcolor=gray!10, 
  linewidth=0.5pt,         
  frametitlerule=true,     
  roundcorner=5pt          
]{definition}{Definition}

\newmdtheoremenv[
  backgroundcolor=gray!10, 
  linewidth=0.5pt,         
  frametitlerule=true,     
  roundcorner=5pt          
]{theorem}{Theorem}

\newcommand{\opr}{{\ensuremath{\sf{\mathcal Vrf}}}\xspace}
\newcommand{\dev}{{\ensuremath{\sf{\mathcal Prv}}}\xspace}
\newcommand{\prv}{{\ensuremath{\sf{\mathcal Prv}}}\xspace}
\newcommand{\vrf}{{\ensuremath{\sf{\mathcal Vrf}}}\xspace}
\newcommand{\f}{{\ensuremath{\sf{\mathcal F}}}\xspace}
\newcommand{\s}{{\ensuremath{\sf{\mathcal S}}}\xspace}
\newcommand{\inp}{{\ensuremath{\sf{\mathcal I}}}\xspace}
\newcommand{\outp}{{\ensuremath{\sf{\mathcal O}}}\xspace}
\newcommand{\chal}{{\ensuremath{\sf{\mathcal Chal}}}\xspace}

\newcommand{\acron}{\ensuremath{\sf{SLAPP}}\xspace}
\newcommand{\posx}{\ensuremath{\sf{PoSX}}\xspace}
\newcommand{\pox}{\ensuremath{\sf{PoX}}\xspace}

\newcommand{\adv}{{\ensuremath{\sf{\mathcal Adv}}}\xspace}
\newcommand{\devprivkey}{{\ensuremath{\sf{sk_{\dev}}}}\xspace}
\newcommand{\devpubkey}{{\ensuremath{\sf{pk_{\dev}}}}\xspace}
\newcommand{\vrfprivkey}{{\ensuremath{\sf{sk_{\vrf}}}}\xspace}
\newcommand{\vrfpubkey}{{\ensuremath{\sf{pk_{\vrf}}}}\xspace}
\newcommand{\ignore}[1]{}

\newcommand{\revision}[1]{{\color{black}{#1}}}

\newcounter{protocol}
\newenvironment{protocol}[1]
  {\par\addvspace{\topsep}
   \noindent
   \tabularx{\linewidth}{@{} X @{}}
    \hline
    \refstepcounter{protocol}\textbf{Protocol \theprotocol:} #1 \\
    \hline}
  { \\
    \hline
   \endtabularx
   \par\addvspace{\topsep}}

\newcommand*\emptycirc[1][.8ex]{\tikz\draw (0,0) circle (#1);} 
\newcommand*\halfcirc[1][.8ex]{%
  \begin{tikzpicture}
  \draw[fill] (0,0)-- (90:#1) arc (90:270:#1) -- cycle ;
  \draw (0,0) circle (#1);
  \end{tikzpicture}}
\newcommand*\fullcirc[1][.8ex]{\tikz\fill (0,0) circle (#1);} 
\begin{document}
%


\title{Poisoning Prevention in Federated Learning and Differential Privacy via Stateful Proofs of Execution}

\author{
\IEEEauthorblockN{Norrathep Rattanavipanon\IEEEauthorrefmark{1} and Ivan De Oliveira Nunes\IEEEauthorrefmark{2}}~\\
\IEEEauthorblockA{\IEEEauthorrefmark{1}College of Computing, Prince of Songkla University}~\\
\IEEEauthorblockA{\IEEEauthorrefmark{2}Department of Informatics, University of Zurich}
\vspace{-.2in}
}

\maketitle

\begin{abstract}

The rise in IoT-driven distributed data analytics, coupled with increasing privacy concerns, has led to a demand for effective privacy-preserving and federated data collection/model training mechanisms. In response, approaches such as Federated Learning (FL) and Local Differential Privacy (LDP) have been proposed and attracted much attention over the past few years.
However, they still share the common limitation of being vulnerable to poisoning attacks wherein adversaries compromising edge devices feed forged (a.k.a. ``poisoned") data to aggregation back-ends, undermining the integrity of FL/LDP results.

In this work, we propose a system-level approach to remedy this issue based on a novel security notion of Proofs of Stateful Execution (\posx) for IoT/embedded devices' software. To realize the \posx concept, we design \acron: a \underline{S}ystem-\underline{L}evel \underline{A}pproach for \underline{P}oisoning \underline{P}revention. \acron leverages commodity security features of embedded devices -- in particular ARM TrustZone-M security extensions -- to verifiably bind raw sensed data to their correct usage as part of FL/LDP edge device routines. As a consequence, it offers robust security guarantees against poisoning. Our evaluation, based on real-world prototypes featuring multiple cryptographic primitives and data collection schemes, showcases \acron's security and low overhead.

\end{abstract}

\maketitle

\section{Introduction}
\label{sec:intro}
With the rise of IoT and distributed big data analytics, data produced by edge devices have become increasingly important to understand users' behaviors, enhance the user experience, and improve the quality of service. At the same time, privacy concerns have scaled significantly fueled by the collection (or leakage) of sensitive user data~\cite{cambridgeanalytica,GDPR1,GDPR2,GDPR3,viceroy}.
To reconcile privacy and utility, several mechanisms have been proposed to enable efficient and privacy-preserving collection of data produced by (typically resource-constrained) edge IoT devices. For example, Google has proposed Federated Learning (FL) aiming to collect and train models based on user data in a distributed, lightweight, and privacy-preserving fashion~\cite{konevcny2016federated}; \revision{Microsoft employs a mechanism based on Local Differential Privacy (LDP) to collect statistics of sensitive telemetry data across millions of devices~\cite{ding2017collecting}}.

\begin{figure}
  \centering
    \includegraphics[width=\columnwidth]{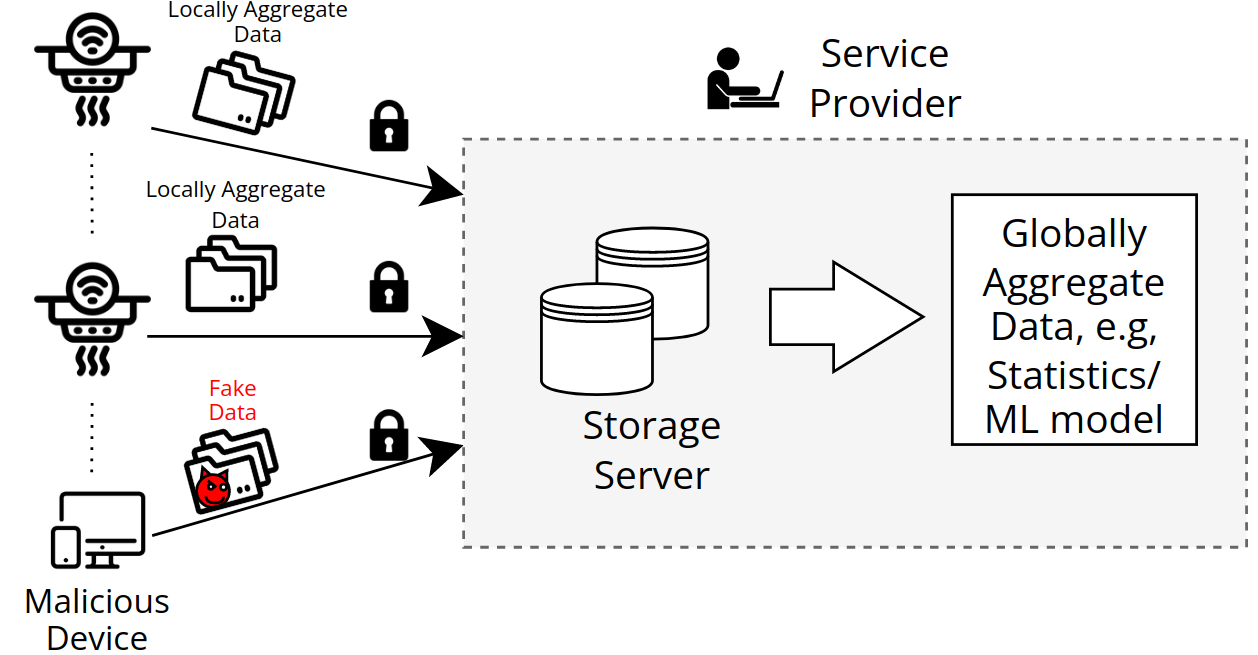}
    \caption{Poisoning in FL/LDP-based systems}\label{fig:system}
    \vspace{-1em}
\end{figure}

As illustrated in Fig.~\ref{fig:system}, a crucial security obstacle in the adoption of these schemes is an adversary (\adv) that feeds back-end aggregators with forged data, sabotaging the entire collection process even when the other participants (other edge devices and the back-end) are honest.
These attacks, known as \emph{poisoning attacks}, have been widely recognized and are currently considered a major problem for both FL~\cite{tolpegin2020data} and \revision{LDP-based data collection mechanisms}~\cite{cao2021data}. 
\revision{In particular, embedded/IoT devices are highly susceptible to software exploits that potentially lead to these attacks due to their inherent lack of security mechanisms~\cite{ammar2024sok,tsudik2024staving}. For instance, \adv can exploit memory-safety vulnerabilities such as buffer overflows (e.g., CVE-2020-10023~\cite{cve_2020_10023}) or heap-based exploits (e.g., CVE-2017-14201~\cite{cve_2017_14201}) to gain remote code execution power; this in turn enables poisoning attacks directly on these devices. Moreover, IoT ecosystems often exhibit \emph{monocultures}~\cite{tsudik2024staving}, where the same types of devices (possibly containing the same exploitable vulnerability) are deployed by service providers. 
This intensifies the risk of poisoning attacks as \adv can compromise multiple IoT devices simultaneously (via the same exploit), facilitating poisoning on a large scale.
}

Existing mitigation techniques are either data-driven (e.g., in the case of data poisoning detection~\cite{cao2021data,cao2019understanding, li2023fine,fang2020local}) or algorithmic, e.g., by making FL and LDP mechanisms more resilient against these attacks~\cite{blanchard2017machine,guerraoui2018hidden, feng2014robust, jagielski2018manipulating,sun2021fl}. In both cases, security is by design best-effort, since these approaches cannot detect/prevent data poisoning at its source, i.e., at the edge devices themselves (typically, resource-constrained IoT devices).

In this work, we propose a systematic and practical treatment to address data poisoning at its source by leveraging architectural security features of contemporary embedded systems. This ensures that raw sensed data is correctly linked with its respective local processing, ultimately generating local aggregated data whose integrity and authenticity can be verified by back-ends in \revision{FL/LDP-based} applications.

Specifically, we build upon the recently introduced concept of Proofs of Execution (\pox)~\cite{apex} for simple embedded systems. \pox allows a low-end embedded device -- Prover (\dev) -- to convince a remote Verifier (\vrf) that a specific function \f has been executed successfully (i.e., from its first to its last instruction) on \dev. Furthermore, PoX binds obtained results (or outputs) to a timely instance of this execution. A similar notion~\cite{flicker} has also been explored in high-end systems (e.g., general-purpose computers and servers, as opposed to embedded devices) based on trusted platform modules (TPMs)~\cite{tpm}.

Intuitively, \pox can convey if the result received by an FL/LDP back-end (local aggregate data) truly originates from an edge device that has obtained this data through the correct execution of the expected software, thus thwarting poisoning attacks. 
On the other hand, as detailed next, the adoption of \pox in \revision{FL/LDP-based mechanisms} introduces unique non-trivial challenges that require re-thinking and re-designing existing \pox methods for this purpose.

\ignore{
\begin{figure}
  \centering
    \includegraphics[width=\columnwidth]{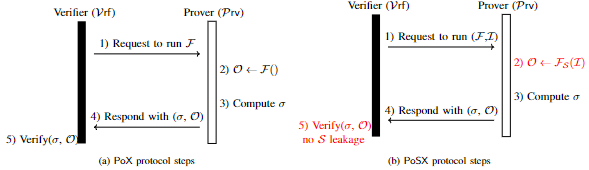}
    \caption{Standard \pox vs \posx}\label{fig:pox}
    \vspace{-1em}
\end{figure}
}

\subsection{On the Insufficiency of Classic \pox to Avert Poisoning}

\pox is a challenge-response protocol composed of the following steps:
\begin{enumerate}
    \item \vrf sends an authenticated request containing a cryptographic challenge (\chal) and asking \dev to execute \f.
    \item \dev authenticates the request, executes $\f$ obtaining output \outp, and generates a cryptographic proof $\sigma$ of this execution by measuring (signing or a MAC-ing) \f's implementation in program memory along with received \chal, produced \outp, and execution metadata that conveys to \vrf if \f execution was performed correctly.
    \item \dev returns the output and proof ($\sigma$, \outp) to \vrf.
    \item \vrf verifies whether $\sigma$ corresponds to the expected \f code, received output \outp, and expected execution metadata. If so, it concludes that \f has executed successfully on \dev with result \outp.
\end{enumerate}

Step 2 above must be securely implemented by a root of trust (RoT) within \dev to ensure (1) temporal consistency between \f's measurement and its execution, (2) correctness of \f execution and generated \outp at run-time, and (3) confidentiality of the cryptographic secret used to compute $\sigma$. This RoT implementation must be unmodifiable, even when \dev's application software is fully compromised. The latter is typically obtained through hardware support, e.g., from ARM TrustZone~\cite{trustzone} (see Section~\ref{sec:bg_tz}) or similar mechanisms.

We observe that the aforementioned \pox notion has important practical limitations. It assumes \f to be: (1) inputless, i.e., \f cannot depend on inputs external to \prv, and (2) stateless, i.e., \f must not depend on states produced by prior \pox instances in \prv.
As a result, PoX is only suitable for simple self-contained programs that may process locally collected data (via \prv local I/O interfaces) but do not depend on external inputs or prior execution states. This assumption becomes problematic when attempting to apply \pox to \revision{FL/LDP-based mechanisms}.

For FL integrity, \dev should prove that a training function \f  was executed on local training dataset $D$ using \vrf-supplied global weights $W$ and training parameters.
Moreover, no portion of $D$ should be revealed to \vrf, requiring multiple \pox instances (e.g., multiple sensing routines executed over time) to correctly produce all data points in $D$. 
As detailed in Section~\ref{sec:bg_fl/ldp}, similar requirements exist in \revision{LDP algorithms}. Therefore, standard \pox can not be applied in these settings.

\subsection{Our Contributions}

To address the aforementioned limitations, this work introduces the security notion of Proof of Stateful Execution (\posx) to enable \textbf{input validation} and \textbf{state preservation}, in addition to classic \pox guarantees. The former relaxes the constraint of inputless functions in traditional \pox, while the latter ensures that \posx can use \dev pre-existing states as long as they originate from a prior authentic \posx execution.
In essence, PoSX offers assurance that execution of \f, computed with authentic input \inp and state \s, denoted $\f_\s(\inp)$, occurred faithfully, without disclosing \s to \vrf.

To realize \posx on real-world IoT settings, we design and implement \acron: a \underline{S}ystem-\underline{L}evel \underline{A}pproach for \underline{P}oisoning \underline{P}revention.
\acron's design and security rely on the commercially available TrustZone-M Security Extension, widely present even in low-end embedded devices: those based on ARM Cortex-M Micro-Controller Units (MCUs).
This facilitates the immediate real-world implementation of \acron onto current IoT devices.

We show that \acron can support a wide range of data collection schemes, including FL and LDP, all with poisoning-free guarantees.
Compared to prior data-centric mitigations in FL~\cite{sun2021fl, cao2019understanding, ma2022shieldfl} or LDP~\cite{cao2021data}, 
\acron offers two key benefits: 
\revision{as a system-level approach, it is agnostic to the underlying collection scheme (and implementation thereof), thereby capable of supporting both FL and LDP without changes in its trusted computing base (TCB), i.e., its TCB remains the same for any function \f. 
Secondly, it primarily operates on the \dev-side while requiring one additional verification operation on \vrf. This makes \acron complementary to many server-side techniques, allowing seamless integration and the ability to further benefit from these techniques. 
We elaborate more on these points in Section~\ref{sec:acron-with-others}.}
In summary, our anticipated contributions are the following:

\begin{enumerate}
    \item {\bf New \posx Security Notion:} we define a new security primitive, called Proof of Stateful Execution (\posx). \posx retains the same guarantees as classic \pox while addressing its limitations of inputless and stateless execution and maintaining privacy of underlying execution states {\it vis-a-vis} \vrf.

    \item {\bf Practical Poisoning Prevention:} We develop \acron:  a design to realize \posx that is applicable to resource-constrained embedded devices. We integrate \acron with FL and LDP implementations to support poison-free instantiations of these algorithms without loss of privacy.

    \item {\bf Real-World Prototypes:} To validate \acron and foster reproducibility, we provide three implementation variants, each utilizing distinct cryptographic schemes: symmetric, traditional asymmetric, and quantum-resistant primitives. These implementations are prototyped and opensourced~\cite{repo} on a real-world IoT development board: NUCLEO-L552ZE-Q~\cite{nucleo}. 

    \item {\bf Evaluation:}
    We conduct various experiments to assess \acron's efficiency. 
    Our results demonstrate small runtime and memory overhead atop the baseline, in exchange for increased security and flexibility. Finally, we provide detailed case studies highlighting \acron's efficiency and efficacy in thwarting poisoning attacks within \revision{FL/LDP-based mechanisms}.
\end{enumerate}


\section{Background}

\subsection{Privacy-Preserving Data Collection  Schemes}\label{sec:bg_fl/ldp}

In this section, we overview FL and LDP.
We use the notation $\phi$ to indicate an empty/null variable.~\\ 

\textbf{Local Differential Privacy-based Data Collection (\texttt{LDP-DC}).}
The goal of this scheme is for an IoT sensor, \dev, to output a noisy sensor value $\outp$ such that it preserves $\epsilon$-LDP; informally speaking, $\epsilon$-LDP guarantees that $\outp$ leaks no information about its original value $\outp'$ except with a small probability constrained by $\epsilon$.
At the same time, collecting $\outp$ from a number of \dev-s allows \vrf to obtain certain statistics of the collected sensor values with high confidence. Various LDP mechanisms have been proposed to achieve $\epsilon$-LDP in different settings. 
As an example case, we focus on the LDP mechanism called Basic RAPPOR~\cite{erlingsson2014rappor} noting that the concepts introduced in this work can be generalized to other LDP-based mechanisms.

\texttt{LDP-DC}, as detailed in Algorithm~\ref{alg:ldp}, describes how \dev performs data collection using Basic RAPPOR. First, \dev collects a sensor reading $\outp'$ by calling the sensor function \f on a given input \inp, i.e., $\f(\inp)$. Assuming $\outp'$ can be represented using $k$-bit unsigned integer, \dev next performs a unary encoding ($UE$) that transforms $\outp'$ into a $2^k$-bit vector $b$ in which only the $\outp^{th}$ bit is set to 1 and other bits are 0. 

\begin{algorithm}
    \footnotesize
    \caption{Implementation of \texttt{LDP-DC} on \dev}\label{alg:ldp}
    \DontPrintSemicolon
    \SetKwInOut{Input}{Input}
    \SetKwInOut{Output}{Output}
    \SetKwInOut{State}{State}
    \Input{Sensor function \f, Input to sensor function \inp, Basic RAPPOR parameters $f$, $p$ and $q$}
    \Output{Noisy sensor output $\outp$}
    \State{$PRR$ mapping $B$}
    
    \SetKwProg{Fn}{Func}{:}{}
    \SetKwFunction{Func}{\texttt{LDP-DC}}
    \Fn{\Func{$\f$, $\inp$ $f$, $p$, $q$}}{
        $\outp' \leftarrow \f(\inp)$\;
        $b \leftarrow UE(\outp')$\;
        $b' \leftarrow PRR(b, f, B)$\;
        $\outp \leftarrow IRR(b', p, q)$\;
        \KwRet~\outp\;
    }
\end{algorithm}

Next, on input $b$ and parameter $f$, \dev invokes Permanent Randomized Response ($PRR$) function and produces a $2^k$-bit noisy vector $b'$ as output, where $b'_i$ -- the $i^{th}$ bit of $b'$ -- is computed as:
\begin{equation}
\footnotesize
    b'_i = 
    \begin{cases}
        1,  & \text{with probability } \frac{f}{2} \\
        0,  & \text{with probability } \frac{f}{2} \\
        b_i,  & \text{with probability } 1-f \\
    \end{cases}
\end{equation}

Once $b'$ is generated, \dev caches an input-output mapping ($b$, $b'$) to a local state variable $B$ (i.e., $B[b] = b'$) and, for all future encounters of the same input $b$, returns $B[b]$ without recomputing the entire $PRR$ function.

Finally, on input $b'$ and parameters $p$ and $q$, \dev executes Instantaneous Randomized Response ($IRR$) function that returns a $2^k$-bit binary vector $\outp$ to \vrf such that:
	
\begin{equation}
\footnotesize
    \mathbb{P}(\outp_i=1) =
    \begin{cases}
        p, & \text{if } b'_i=1 \\
        q, & \text{otherwise}
    \end{cases}
\end{equation}

Using \texttt{LDP-DC}, \vrf can estimate $\tilde{f}_{x}$, the frequency of sensor value $x$ by:
\begin{equation}
\footnotesize
	\tilde{f}_{x} = \dfrac{c_{x} - (q+\frac{1}{2}fp-\frac{1}{2}fq)n}{(1-f)(p-q)n}
\end{equation}

where $c_{x}$ represents the number of reports that have $x^{th}$ bit set and $n$ is the number of reports received by \vrf. We refer the interested reader to the Basic RAPPOR paper~\cite{erlingsson2014rappor} for details on how to select $f,p,q$ to satisfy $\epsilon$-LDP.

\revision{We emphasize that the state variable $B$ serves as a critical component to the privacy of this scheme.
To ensure $\epsilon$-LDP, $B$ must be accurately updated in the current Basic RAPPOR session and its updated value must be carried forward to the subsequent session.
Moreover, $B$ must be oblivious to \vrf; otherwise, \vrf can reverse the $PRR$ operation to recover the original sensor reading $\outp'$.
Also, this scheme requires input arguments $f$, $p$, and $q$ as part of its execution.}
As noted earlier, however, standard PoX does not support execution using external input arguments or \dev local states that must be oblivious to \vrf.~\\

\begin{algorithm}
    \footnotesize
    \caption{Implementation of \texttt{FL-DC} on \dev}\label{alg:fl}
    \DontPrintSemicolon
    \SetKwProg{Fn}{Function}{:}{}
    \SetKwInOut{Input}{Input}
    \SetKwInOut{Output}{Output}
    \SetKwInOut{State}{State}
    \Input{Sensor function \f, Input to sensor function \inp}
    \Output{$\phi$}
    \State{Local dataset $D$}
    
    \SetKwProg{Fn}{Func}{:}{}
    \SetKwFunction{Func}{\texttt{Sense-Store}}
    \Fn{\Func{\f, \inp}}{
        $\outp \leftarrow \f(\inp)$\;
        $D$.append(\outp)\;
    }
    \vspace{2mm}
    
    \Input{Globally trained weights $W$, number of epochs $t$, learning rate $\alpha$}
    \Output{Locally trained weights $\outp$ }
    \State{$D$}
    \SetKwProg{Fn}{Func}{:}{}
    \SetKwFunction{Func}{\texttt{Train}}
    \Fn{\Func{$W$, $t$, $\alpha$}}{
        \For{$k \gets 1$ \KwTo $t$}{
            \tcp{$\nabla$ is a gradient function}
            $W \leftarrow W - \alpha \cdot \nabla(W; D)$\;
        }
        $\outp \leftarrow W$\;
        \KwRet~\outp\;
    }
    
\end{algorithm}

\textbf{Federated Learning-based data collection (\texttt{FL-DC}).}
Contrary to \texttt{LDP-DC}, \texttt{FL-DC}~\cite{konevcny2016federated} requires \dev to send a locally trained machine learning (ML) partial model to \vrf instead of sensor readings. This keeps raw sensor data local to \prv and not directly accessible by \vrf.

A typical \texttt{FL-DC} consists of two phases depicted in Algorithm~\ref{alg:fl}.
In the first phase, \dev invokes the \texttt{Sense-Store} function to collect a raw sensor reading and store it in a local list $D$.
This phase can be repeated multiple times over a certain period to collect more training data on \dev.
Once sufficient training data is gathered, 
\dev receives globally trained weights $W$ from \vrf along with the training parameters ($t$ and $\alpha$).
It then triggers the \texttt{Train} function that utilizes $W$ as a base model to train the data in $D$.
This function outputs the locally trained weights $\outp$ without disclosing $D$ to \vrf.

After collecting weights $\outp$-s from multiple devices, \vrf can aggregate them using several methods, e.g., FedAvg~\cite{konevcny2016federated} averages the received weights and sets the result as the new global weights: $W \leftarrow W + \eta\cdot\sum_{i=1}^{m} (\outp_{i} - W) / m$ for some global learning rate $\eta$.

\revision{Similar to \texttt{LDP-DC}, executing \texttt{FL-DC} functions relies on both input arguments $W$, $t$ and $\alpha$ as well as \dev-local state $D$.} 
Consequently, \texttt{FL-DC} integration with classic \pox faces the same challenges as \texttt{LDP-DC}.

\subsection{ARM TrustZone-M Security Extensions}\label{sec:bg_tz}

ARM TrustZone-M is a hardware security extension that enables a trusted execution environment (TEE) in ARM Cortex-M MCUs commonly used in low-cost and energy-efficient IoT applications. TrustZone divides software states into two isolated worlds: \emph{Secure} and \emph{Non-Secure}. The Non-Secure world contains and executes (untrusted) application software while security-critical (trusted) software is stored and runs in the Secure world.

In particular, we leverage two security properties of ARM TrustZone-M in this work:

\begin{itemize}
    \item {\bf Hardware-enforced World Isolation.} TrustZone-M ensures complete isolation of these worlds by implementing several hardware controls (i.e., SAU/IDAU) to enforce access control to hardware resources (e.g., program and data memory, peripherals) for these two worlds.
    With isolation in place, TrustZone ensures the Non-Secure World is unable to access any code and data located in the Secure World. This assures that the Secure world remains secure even if an adversary can fully modify or compromise the Non-Secure world software state.

    \item {\bf Controlled Invocation.} 
    \revision{In TrustZone-M, the only legal way for the Non-Secure world to access a function inside the Secure World is by making a call to predefined entry points. These entry points are located in a designated area within the Secure World, known as the Non-Secure Callable (NSC) region.
    As a part of the Secure World, the NSC region cannot be tampered with by the Non-Secure World's software. 
    As a result, this mechanism combined with TrustZone-M secure context switching enables controlled invocation of the Secure World functions, preventing attacks that aim to compromise secure functions by executing them partially, i.e., by jumping into or exiting from the middle of the function.
    }
\end{itemize}

\section{System Model and Assumptions}

\subsection{Network and Usage Model}


We consider an IoT setting as shown earlier in Fig.~\ref{fig:system}, consisting of two entity types: one \opr and multiple \dev-s. \dev is a resource-constrained sensor deployed in a physical space of interest, e.g., a smart home, office, or factory. \opr is a remote service provider that orchestrates these \dev-s.
As an example, \dev-s could be smart light bulbs that are used in many smart homes and can be controlled by the end-user through Philips \opr's application services; or Samsung could act as \opr that provides a service for the end-user to command all SmartThings-compatible devices.

Besides offering this service, \opr wishes to collect sensor data generated by \dev-s to further improve the service performance or enhance the user experience. 
In this work, \vrf has the option to employ \texttt{LDP-DC} or \texttt{FL-DC} as its preferred data collection scheme, depending on the desired outcome and privacy considerations.
As poisoning attacks could sabotage the collection outcome, \vrf also aims to detect and prevent such attacks to safeguard the authenticity of the outcomes.

\subsection{Adversary Model}


We consider an \adv who can modify/compromise \dev's application software at will.
Once compromised, \adv can access, modify, or erase any code or data in \dev unless explicitly protected by hardware-enforced access control rules.
Consequently, \adv may use this ability to perform poisoning attacks by corrupting a sensor function \f or its execution to spoof arbitrary results sent to \opr.
\revision{In the context of FL, in addition to data poisoning, \adv may use this capability to launch \textit{model} poisoning attacks that aim to compromise the global machine learning model by introducing malicious local models (i.e., gradient updates) to \vrf.}
Invasive hardware-based/physical attacks (e.g., fault-injection attacks or physical hardware manipulation) are out of scope in this work, as they require orthogonal tamper-proofing techniques~\cite{ravi2004tamper}.

We also assume \adv has full control over the communication channel between \dev and \opr.
They may perform any network-based attacks, e.g., reading, modifying, replaying, or dropping any message sent from/to \dev. 

Further, we consider \adv to be adaptive~\cite{tramer2020adaptive}, i.e., it is aware of the algorithm and all the specifics of the protocol executed between \vrf and \prv. As a result, \adv is allowed to modify its attack strategy by modifying \dev's software state (except for hardware-enforced protections) and the communication between \vrf and \prv to attempt to circumvent the proposed defense.

Finally, in line with the standard LDP and FL \adv models, we also consider the possibility of malicious \vrf.
In the latter, \adv's goal is to learn sensitive data on \dev while executing the protocol. 


\subsection{Device Model}

\prv-s are small embedded/IoT devices equipped with TrustZone-M, e.g., ARM embedded devices running on Cortex-M23/33 MCUs, which are optimized for low-cost and energy efficiency. 
%
Following the PoX assumption~\cite{apex}, we assume that the function whose execution is being proven (\f, i.e., the code implementing the data collection task according to the underlying scheme) is correct and contains no implementation bug that can lead to run-time exploits within itself. In practice, \opr can employ various pre-deployment vulnerability detection techniques to fulfill this requirement~\cite{celik2018soteria}.

We also adhere to standard TEE-based security assumptions, i.e.,  we assume the small TCB implementing the \posx RoT located inside TrustZone's Secure World is trusted and TrustZone hardware modules are implemented correctly such that \adv cannot modify this RoT implementation or violate any security guarantees implemented by the Secure World-resident code. 
%
%
The latter implies the existence of secure persistent storage, exclusively accessible by the Secure World and unmodifiable when the device is offline.
This storage is used to store our TCB along with a counter-based challenge $c$ and two cryptographic keys: \devprivkey and \vrfpubkey, where \devprivkey corresponds to \dev private key, whose public counterpart \devpubkey is known to \opr. Similarly, \vrfpubkey denotes \vrf public key with its private counterpart \devprivkey securely managed by \vrf.
In TrustZone-M implementations, secure persistent memory is supported by a standard secure boot architecture~\cite{arbaugh1997secure} with the physical memory storing cryptographic keys being physically inaccessible through I/O interfaces (e.g., USB/J-TAG, etc.).
 
Finally, we assume these keys are correctly distributed to \vrf and \dev out-of-band, e.g., by physically fusing these keys on \dev during manufacture time or using any key-provisioning mechanism~\cite{moharana2017secure} after \dev deployment.

\section{\posx and Associated Definitions}\label{sec:pox-def}


To define \posx security goal, we start by revisiting classic \pox guarantees and their limitations. Then we formulate auxiliary notions that address each limitation and in conjunction imply \posx end-to-end goal.

\begin{definition}[PoX Security\label{def:pox}~\cite{apex}]\footnotesize
    Let \f represent an arbitrary software function (code) execution of which is requested by \vrf on \dev, producing output $\outp$.
    A protocol is considered PoX-secure
    if and only if the protocol outputting $\top$ implies:
    \begin{enumerate}[label=(\roman*)]
        \item $\f$ code (as defined by \vrf) executes atomically and completely between $\vrf$ sending a request and receiving the response, and
        \item \outp is a direct outcome of this execution of $\f()$.
    \end{enumerate}
\end{definition}

Definition~\ref{def:pox} states the classic \pox security notion, as described in~\cite{apex}.
A violation of conditions (i) or (ii) in Definition~\ref{def:pox} must be detectable by \vrf , resulting in a protocol abort (i.e., by outputting $\bot$).
As discussed in~\cite{apex}, this notion can be used to construct ``sensors that cannot lie" irrespective of compromised software states.

However, it only supports self-contained IoT applications that are independent of \vrf-defined inputs or pre-computed states due to two limitations:



\textbf{L1 -- Lack of input validation.} Definition~\ref{def:pox} considers \emph{inputless} \f functions. This is because classic \pox does not support verification of the origin and integrity of inputs received by \prv. This limitation is significant for applications where \vrf must provide input \inp as part of \f execution on \dev (e.g., FL/LDP cases discussed in Section~\ref{sec:bg_fl/ldp}).

\textbf{L2 -- No state preservation across \pox instances.}
Definition~\ref{def:pox} only supports \pox of stateless \f functions.
In other words, \f must rely solely on data generated/acquired within its current execution instance and must not depend on \prv states (denoted \s) produced by prior executions.
Similar to {\bf L1} case, a \pox protocol satisfying Definition~\ref{def:pox} provides no guarantee or validation to the correct use of some pre-existent/expected state \s in \prv. Thus, attacks that tamper with \s in between subsequent \pox instances may result in illegal alteration of the end result \outp.
    
\textit{\textbf{Remark:} when \s contains only public information, \textbf{L2} can be obviated by \textbf{L1} by making \s a part of the authenticated output of a \pox instance and used as a \vrf-defined input to \f in a subsequent \pox instance. However, when \s must remain hidden from \vrf (the case of our target applications -- recall Section~\ref{sec:bg_fl/ldp}), consistency of \s must be ensured locally at \prv, making \textbf{L1} and \textbf{L2} independent challenges.}

To address these limitations systematically,
we introduce two new \pox-related security notions.
As these notions may be of independent interest, we first present them separately and finally compose them into an end-to-end \posx goal.


\begin{definition}[IV-\pox Security\label{def:iv-pox}]\footnotesize
    Let \f represent an arbitrary software function (code) and $\inp$ represent an input, both defined by \vrf. Let $\outp$ represent the output and $\sigma$ denote a proof of $\f(\inp)$ execution produced by \dev.
    A protocol is considered IV-\pox-secure
    if and only if the protocol outputting $\top$ implies:
    \begin{enumerate}[label=(\roman*)]
        \item $\f$ executes \underline{with input arguments $\inp$}, atomically and completely between $\vrf$ sending the request and receiving $\sigma$, and
        \item \outp is a direct outcome of this $\f(\inp)$ execution
    \end{enumerate}
\end{definition}

To overcome \textbf{L1}, we present the notion of input-validating \pox (or IV-\pox), as shown in Definition~\ref{def:iv-pox}.
In IV-\pox, \vrf aims to execute \f with its own provided input \inp.
Thus, the proof generated by \dev, $\sigma$, must not only validate atomic and complete execution of \f but also that it was invoked with the correct input requested by \vrf (as captured in condition (i) of Definition~\ref{def:iv-pox}).
Thus, an IV-\pox protocol must ensure \outp authenticity concerning \inp, i.e., \outp is generated by executing \f correctly using the expected input \inp.




\begin{definition}[SP-\pox Security\label{def:sp-pox}]\footnotesize
    Let \f represent an arbitrary software function (code) requested by \vrf to run on \dev with state $\s$, where \s was produced by some prior \pox on \prv but is oblivious to \vrf. Let $\outp$ represent the output and $\sigma$ denote a proof of $\f$ execution produced by \dev.
    A protocol is considered SP-\pox-secure if and only if the protocol outputting $\top$ implies:
    \begin{enumerate}[label=(\roman*)]
        \item $\f$ executes atomically and completely between $\vrf$ sending the request and receiving $\sigma$ with \prv state corresponding to \s when \f execution starts (denote this execution by $\f_\s()$), and
        \item \outp is a direct outcome of this $\f_\s()$ execution, and
        \item \vrf cannot infer the value of \s beyond what is revealed by \outp, and
        \item \s was not modified between the current $\f_\s()$ execution and the prior \vrf-authorized \pox. 
    \end{enumerate}
\end{definition}

To address {\bf L2}, we specify the State-Preserving \pox (SP-\pox) notion in Definition~\ref{def:sp-pox}.
Similar to IV-\pox security, the SP-\pox notion  
specifies the first two conditions to ensure that, in addition to atomic \f execution, $\sigma$ also conveys two critical aspects: (1) correct use of \s during \f execution and (2) dependence of \outp on \f and \s.
In addition, it requires \s privacy vis-a-vis \vrf and prohibits \s modification in between subsequent \pox instances.

Finally, Definition~\ref{def:posx} combines IV-\pox and SP-\pox to state the goal of \posx-Security.

\begin{definition}[\posx Security\label{def:posx}]
\footnotesize
    A scheme is \posx-secure
    if and only if it satisfies both IV-\pox (Def.~\ref{def:iv-pox}) and SP-\pox (Def.~\ref{def:sp-pox}) Security.
\end{definition}


\section{\acron: Realizing \posx with TrustZone-M}\label{sec:posx}


\subsection{Overview of \acron Workflow}

Building on TrustZone-M hardware-enforced world isolation (recall Section~\ref{sec:bg_tz}), our approach is to implement \acron's RoT in and execute it from the Secure World. Meanwhile, normal applications are untrusted (hereby referred to as untrusted software) and reside in the Non-Secure World.

\acron implements three Secure World functions: \texttt{Execute}, \texttt{CheckState}, and \texttt{SetState}.
\texttt{Execute} serves as the main call to execute $\f_\s(\inp)$ on \dev and compute a proof of this stateful execution. \texttt{CheckState} and \texttt{SetState} are used to authenticate state \s used in \texttt{Execute}.

\acron workflow is depicted in Fig.~\ref{fig:pox}.
At a high level, it enforces the following operation sequence upon receiving a \posx request from \vrf.:
\begin{itemize}
    \item Upon being called by the Non-Secure World, \texttt{Execute} authenticates \vrf request in \ding{185}.
    
    \item Following successful authentication, it starts executing $\f_\s(\inp)$ atomically in Non-Secure World, \ding{186}.
    
    \item Before accessing \s in \ding{187}, $\f_\s(\inp)$ must call \texttt{CheckState} in the Secure World.
    \texttt{CheckState} takes one input argument, representing the current \s value.
    Its task is to authenticate this value by matching it against the latest benign \s value from a prior execution stored within the Secure World. 
    
    \item Similarly, after \s is modified in \ding{187}, execution is trapped into the Secure World via the \texttt{SetState} function. \texttt{SetState} accepts one input argument: the new \s value. It is responsible for committing and securely maintaining the latest benign \s value in the Secure World.
    
    \item Once $\f_\s(\inp)$ execution completes, yielding output \outp, the control returns to \texttt{Execute} in \ding{188}. 
    Then, \texttt{Execute} computes $\sigma$ indicating an authenticated proof of this execution in \ding{189} and returns (\outp, $\sigma$) to the Non-Secure World in \ding{190}, which in turn forwards them to \vrf.
\end{itemize}

With (\outp, $\sigma$), \vrf can determine if $\f_\s(\inp)$ was executed in \dev's Non-Secure World, if \outp is an authentic result of this execution, and if \s was preserved since the prior execution. Importantly, \acron assures that $\sigma$ is not computable unless the aforementioned operation sequence is observed.

\begin{figure}
    \centering
    \includegraphics[width=0.8\columnwidth]{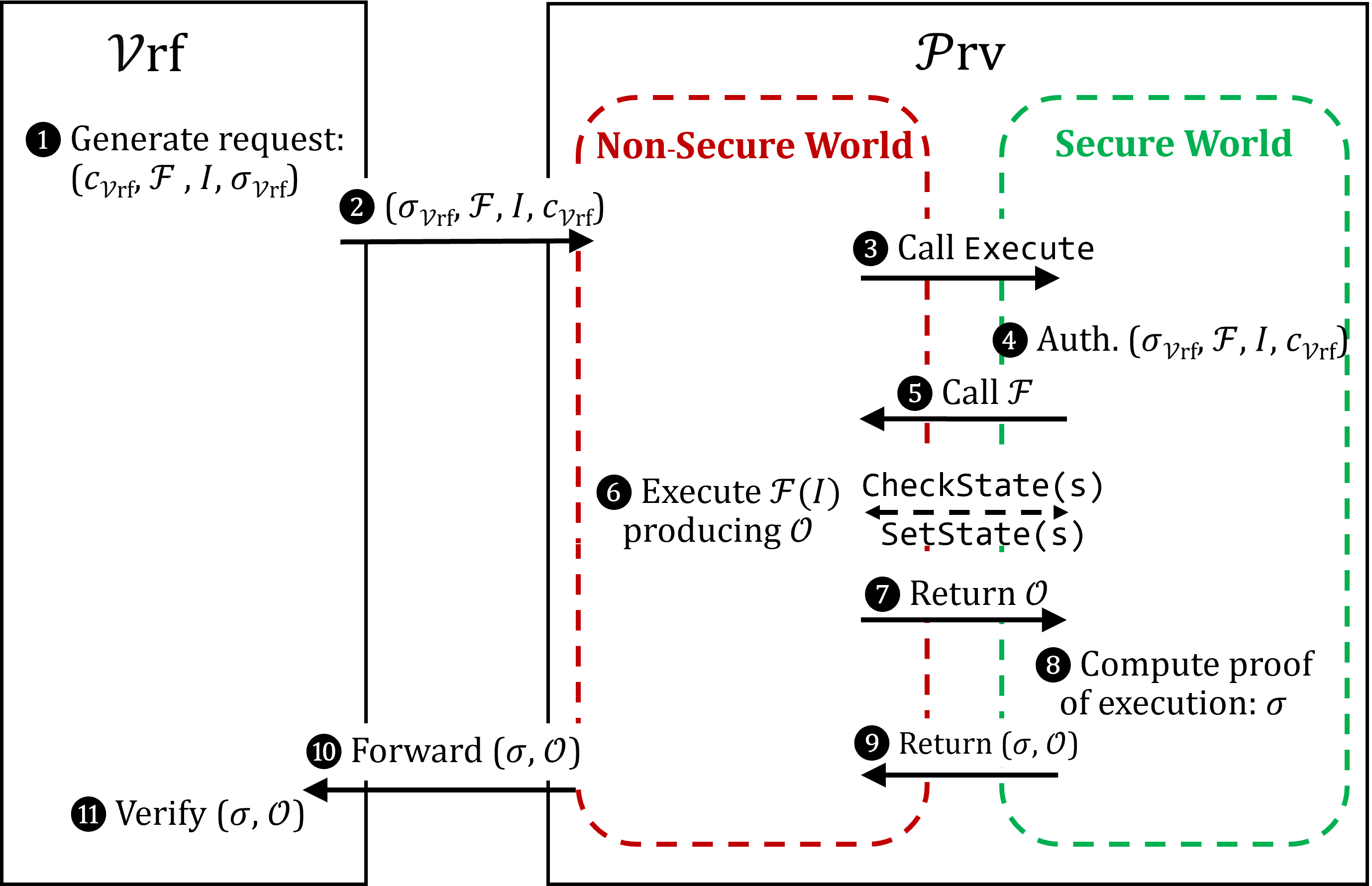}
    \caption{Overview of \acron workflow}
    \label{fig:pox}
    \vspace{-1.5em}
\end{figure} 

\subsection{\acron in Detail}
\label{sec:posx_prot}

Protocol~\ref{prot:posx} details the \vrf $\leftrightarrow$ \dev interation in \acron. 
\begin{center}
    \textbf{Phase 1: Request Generation}
\end{center}
An instance of \acron protocol starts when \vrf generates a \posx request, in Step 1, comprising: (i) an identifier $\f$ for which function to execute; (ii) input arguments \inp; and (iii) a monotonically-increasing counter-based challenge $c_\vrf$. 
This request instructs \dev to execute $\f_\s(\inp)$ in the Non-Secure World on \dev and return the result to \vrf along with proof of this expected execution.
Following this, \vrf signs this request (via \textsf{Sign} function), producing token $\sigma_\vrf$, and attaches this token to the request before sending it to \dev in Step 2. 

\begin{center}
    \textbf{Phase 2: Preparation for $\f_\s(\inp)$ Execution}
\end{center}
In Step 3, untrusted software in \prv must invoke \texttt{Execute} passing the \posx request as a parameter.
In step 4, \texttt{Execute} (i) checks whether the request counter $c_\vrf$ is larger than a local counter $c$ maintained by the Secure World to ensure freshness of the \posx request and prevent replayed attacks; (ii) verifies $\sigma_\vrf$ to confirm authenticity of the request; and (iii) checks that no other \acron instance is active by examining $exec$ flag maintained in the secure world. If any of the checks fail, the process is aborted which implies the inability to produce the end proof $\sigma$. The same applies if \texttt{Execute} is never called.

\begin{protocol}{\textbf{\acron Protocol}\label{prot:posx}} 
    \scriptsize
    \vspace{-5mm}
    \begin{center}
        \textbf{\small\underline{Verifier (\vrf)}}
    \end{center}
    \begin{enumerate}[label={(\arabic*)}]
        \item Generate an authenticated \posx request: 
        \begin{equation*}
            \sigma_\vrf \leftarrow \textsf{Sign}(\vrfprivkey, \textsf{H}(\f, \inp, c_\vrf))
        \end{equation*}
        \item Send ($\sigma_\vrf$, \f, \inp, $c_\vrf$) to \dev
    \end{enumerate}

    \centerline{\hbox to 7cm{\leaders\hbox to 10pt{\hss - \hss}\hfil}\vspace{-2mm}}
    \begin{center}
        \textbf{\small\underline{Prover (\dev)}}
    \end{center}
    \begin{center}
        \textbf{\textit{In Non-Secure World:}}
    \end{center}\vspace{-2mm}
    \begin{enumerate}[label={(\arabic*)}]
        \setcounter{enumi}{2}
        \item Call \texttt{Execute} in Secure World with the received request
    \end{enumerate}\vspace{1mm}

    \begin{center}
        \textbf{\textit{In \texttt{Execute} function, Secure World:}}
    \end{center}\vspace{-2mm}
    \begin{enumerate}[label={(\arabic*)}]
        \setcounter{enumi}{3}
        \item Authenticate the request and abort if $r = \bot$ or $exec = \top$:
        \begin{equation*}
            r \leftarrow (c_\vrf > c) \land \textsf{Verify}(\vrfpubkey, \sigma_\vrf, \f, \inp, c_\vrf)
        \end{equation*}
        \item Update counter: $c \leftarrow c_\vrf$ and initialize: $exec \leftarrow \top$, $stateUsed \leftarrow \bot$ $stateChecked \leftarrow \bot$
        \item Disable interrupts and measure $PMEM$ and \vrf request:
        \begin{equation*}
            h \leftarrow \textsf{H}(PMEM, \f, \inp, c_\vrf)
        \end{equation*}
        \item Call $\f_\s(\inp)$ in Non-Secure World
    \end{enumerate}

    \begin{center}
        \textbf{\textit{In \f function, Non-Secure World:}}
    \end{center}\vspace{-2mm}
    \begin{enumerate}[label={(\arabic*)}]
        \setcounter{enumi}{7}
        \item Run $\f_\s(\inp)$. Pass control to the Secure World when \texttt{CheckState($s$)} is called
    \end{enumerate}

    \begin{center}
        \textbf{\textit{In \texttt{CheckState} function, Secure World:}}
    \end{center}\vspace{-2mm}
    \begin{enumerate}[label={(\arabic*)}]
        \setcounter{enumi}{8}
        \item Perform an integrity check on $s$, store the result to $p$ and return to Non-Secure World:
        $stateChecked \leftarrow (\textsf{H}(s) \stackrel{?}{=} \s_{sec})$. Also, set $stateUsed \leftarrow \top$.
    \end{enumerate}

    \begin{center}
        \textbf{\textit{In \f function, Non-Secure World:}}
    \end{center}\vspace{-2mm}
    \begin{enumerate}[label={(\arabic*)}]
        \setcounter{enumi}{9}
        \item Continue with $\f_\s(\inp)$ execution. Pass control to the Secure World when \texttt{SetState($s$)} is called 
    \end{enumerate}

    \begin{center}
        \textbf{\textit{In \texttt{SetState} function, Secure World:}}
    \end{center}\vspace{-2mm}
    \begin{enumerate}[label={(\arabic*)}]
        \setcounter{enumi}{10}
        \item Securely set $\s_{sec}$ based on input $s$ and return to Non-Secure World:
        \begin{equation*}
            \s_{sec} \leftarrow \textsf{H}(s) \text{ if } (stateChecked \land exec)
        \end{equation*}
    \end{enumerate}
    \normalcolor\vspace{-2mm}
    \begin{center}
        \textbf{\textit{In \f function, Non-Secure World:}}
    \end{center}\vspace{-2mm}
    \begin{enumerate}[label={(\arabic*)}]
        \setcounter{enumi}{11}
        \item Continue with $\f_\s(\inp)$ until the execution is completed, producing output \outp, and then return to its caller, \texttt{Execute}, with \outp
    \end{enumerate}

    \begin{center}
        \textbf{\textit{In \texttt{Execute} function, Secure World:}}
    \end{center}\vspace{-2mm}
    \begin{enumerate}[label={(\arabic*)}]
        \setcounter{enumi}{12}
        \item Abort if $(exec~\land stateUsed~\land \neg stateChecked)$. Otherwise, include \outp to the measurement and compute the proof: 
        \begin{equation*}
            \sigma \leftarrow \textsf{Sign}(\devprivkey, \textsf{H}(h, \outp))
        \end{equation*}
        \item Reset all flags: $exec \leftarrow \bot, stateChecked \leftarrow \bot, stateUsed \leftarrow \bot$
        \normalcolor
        \item Enable interrupts and return with $(\outp, \sigma)$
    \end{enumerate}

    \begin{center}
        \textbf{\textit{In Non-Secure World:}}
    \end{center}\vspace{-2mm}
    \begin{enumerate}[label={(\arabic*)}]
        \setcounter{enumi}{15}
        \item Forward $(\outp, \sigma)$ to \vrf
    \end{enumerate}\vspace{1mm}

    \centerline{\hbox to 7cm{\leaders\hbox to 10pt{\hss - \hss}\hfil}\vspace{-2mm}}

    \begin{center}
        \textbf{\small \underline{Verifier (\vrf)}}
    \end{center}
    \begin{enumerate}[label={(\arabic*)}]
        \setcounter{enumi}{16}
        \item Increment $c_\vrf$ and validate $\sigma$ by:
        \begin{equation*}
            r \leftarrow \textsf{ValidatePoSX}(\devpubkey, \sigma, PMEM', \f,\inp, c_\vrf, \outp)
        \end{equation*}
        The protocol outputs $r$.
    \end{enumerate}
    
\end{protocol}

If all checks succeed, \texttt{Execute} updates the local counter with $c_\vrf$ and initializes three Secure World flags in Step 5:
\begin{itemize}
    \item $exec$ is set to $\top$, indicating an active \acron instance.
    \item $stateChecked$ is set to $\bot$, indicating the status of \s authenticity during \f execution. 
    \item $stateUsed$ is set to $\bot$, indicating the status of \s access during \f execution.
\end{itemize}

In Step 6, \texttt{Execute} prepares \dev for the upcoming execution of \f by disabling all interrupts and taking a snapshot ($h$) as a hash digest reflecting the states of the Non-Secure World's binary code in program memory ($PMEM$) and the parameters received in the \posx request.
It then calls $\f$ on input \inp in the Non-Secure World (Step 7).

To leverage \acron, the implementation of \f must: 
\begin{itemize}
    \item[\textbf{B1:}] At the start of its execution, validate relevant state \s authenticity by calling \texttt{CheckState}. 
    \item[\textbf{B2:}] At the end of its execution, commit the latest \s value to the Secure World by invoking \texttt{SetState}.
    \item[\textbf{B3:}] Throughout its execution, \f must never enable interrupts.
\end{itemize}

\begin{center}
    \textbf{Phase 3: $\f_\s(\inp)$ Execution}
\end{center}

From {\bf B1}, it follows that a call to \f, in Step 8, triggers \texttt{CheckState} of current \s.
In Step 9, \texttt{CheckState}, executed in the Secure World, sets $stateUsed$ to $\top$ and proceeds to verify authenticity of the current \s value, $s$, by comparing $s$ with the latest benign \s value, $\s_{sec}$, stored in the Secure World.
Only if they match, \texttt{CheckState} ascertains $s$ authenticity setting $stateChecked$ to $\top$. Execution of \f is then resumed.
From {\bf B2}, \texttt{SetState} is invoked at the end of \f execution to update $\s_{sec}$ with the new state $s$ if and only if ($stateChecked = exec = \top$).
To optimize storage, especially when $|s|$ is large,
\texttt{SetState} can update $\s_{sec}$ with a hash of $s$, as shown in Step 11. 
This storage optimization comes with the expense of additional runtime overhead for hash computations in \texttt{CheckState} and \texttt{SetState}. We discuss this time-space trade-off further in Section~\ref{sec:ext}.

\f execution completes producing the output \outp and handing the control to \texttt{Execute} with \outp as an input in Step 12.

\begin{center}
    \textbf{Phase 4: Proof Generation}
\end{center}
\texttt{Execute} examines $exec$, $stateUsed$, and $stateChecked$ flags to determine the occurrence of the \texttt{CheckState} $\rightarrow$ \texttt{SetState} sequence.
If this sequence is maintained during \f execution, \texttt{Execute} proceeds to compute the proof $\sigma$ in Step 13 by signing $h$ and \outp using \dev's private key $\devprivkey$. \texttt{Execute} resets all Secure World flags before returning to the Non-Secure World with \outp and $\sigma$, in Step 15.
They are then transmitted to \vrf, in Step 16. 

\begin{center}
    \textbf{Phase 5: Proof Validation}
\end{center}
Upon receiving \outp and $\sigma$, \vrf increments $c_\vrf$ and performs the \posx verification in Step 17 by:
\begin{itemize}
    \item Checking validity of $\sigma$. As \vrf possesses the expected binary of \dev's Non-Secure World, $PMEM'$, this verification involves checking $\sigma$ against $PMEM'$, \f, \inp, $c_\vrf$ and \outp using \dev's public key $\devpubkey$, i.e.: 
    \begin{center}
        $\textsf{Verify}(\devpubkey, \sigma, PMEM', \f, \inp, \outp, c_\vrf) \stackrel{?}{=} \top$
    \end{center}

    \item Inspect $\f$ binary to ensure that it adheres to the expected behaviors: \textbf{B1}, \textbf{B2} and \textbf{B3}.
\end{itemize}

Finally, \acron protocol is considered successful and thus outputs $\top$ if it passes both checks; it aborts with $\bot$ otherwise.

\section{\posx Security Analysis}\label{sec:sec_analysis}

Our security argument is based on the following properties:

\noindent\textbf{P1 - Request Verification.}
In \acron, \dev's TCB always verifies freshness and authenticity of a \posx request before executing \f and generating the proof $\sigma$. This prevents \adv from exploiting forged or replayed requests to manipulate the protocol outcome. See step 4 of Protocol~\ref{prot:posx}.

\noindent\textbf{P2 - Input Validation.} 
A valid $\sigma$ serves as authentication for the correct usage of \inp during \f execution.
This prevents \adv from feeding malicious input to \f while still succeeding in the \acron protocol.
Since \inp is included in a \posx request, this is implied by \textbf{P1} and the fact that \inp is directly used by \acron to invoke $\f$ in step 7 of Protocol~\ref{prot:posx}.

\noindent\textbf{P3 - State Privacy.}
\acron protects privacy of \s from \vrf since the only information \vrf receives are \outp and $\sigma$. 
$\sigma$ is not a function of \s; thus it leaks nothing about \s to \vrf.
Thus, \acron incurs no leakage other than \outp itself.

\noindent\textbf{P4 - State Authenticity.}
In \acron, the successful completion of a protocol instance guarantees the \s value, stored in the Non-Secure World, 
is authentic, i.e., it can only be modified by \vrf-approved software during a protocol instance and
remains unchanged between consecutive instances.
This assurance comes from two observations:

First, $\s_{sec}$ always corresponds to the latest benign \s value because: (1) $\s_{sec}$ cannot be updated outside a protocol instance due to the check of $exec$ flag in Step 11; and (2) $\s_{sec}$ cannot be influenced by forged or replayed \posx requests since the request is always authenticated in Step 4 before $exec$ can be set in Step 5.

Second, any unauthorized modification to \s outside a protocol instance is detected in the subsequent instance by \texttt{CheckState} due to a mismatch between the \s value and $\s_{sec}$.
As a successful \acron protocol implies a successful check from \texttt{CheckState}, \s must be equal to $\s_{sec}$, and, according to the first observation, must contain the latest authentic \s value.
Moreover, \adv may attempt to tamper with \s while \f is executing.
However, doing so requires modification to $PMEM$, which would result in a mismatch with $PMEM'$ during the proof validation in Step 17.

\noindent\textbf{P5 - Atomic Execution.}
\f execution must occur atomically; otherwise, the protocol must fail. This is required to prevent \adv from interrupting \f execution to tamper with its data and execution flows, influencing the outcome. Step 6 realizes this requirement by disabling all interrupts before $\f$ invocation. We explain how this requirement can be relaxed in Section~\ref{sec:ext}.


\noindent\textbf{P6 - Output Authenticity.}
A successful \acron protocol indicates to \vrf that \outp is authentic and generated by executing \f atomically with the \vrf-specified \inp and the correct state \s.
\acron satisfies this property since \textbf{P1} guarantees that \f is always invoked with authentic \inp and \textbf{P5} enforces \f to execute without interruptions before immediately returning to Secure World with \outp. This leaves no opportunities for untrusted software to interrupt \f execution to tamper with \outp.
\adv attempt to change \outp via \s are prevented by \textbf{P4}.~\\

\noindent\textbf{Security Argument.}
We show that \acron satisfies both Definition~\ref{def:iv-pox} and Definition~\ref{def:sp-pox} implying adherence to Definition~\ref{def:posx}.
Per Definition~\ref{def:iv-pox}, IV-\pox security requires \pox assurance for stateless \f functions that run with input arguments, i.e., when $\s = \phi$ and $\inp \neq \phi$. In this scenario, \textbf{P2} and \textbf{P5} ensure the atomic execution of \f with the authentic input \inp provided by \vrf, satisfying condition (i). \textbf{P6} guarantees that \outp is generated as a result of $\f(\inp)$ execution, fulfilling condition (ii).
With both conditions met, \acron achieves IV-\pox security.
Meanwhile, SV-\pox security in Definition~\ref{def:sp-pox} requires secure \pox of a stateful \f executed without input arguments.
Similar to the previous argument, \textbf{P5} and \textbf{P6} directly address conditions (i) and (ii) of SV-\pox security even when $\inp = \phi$.
As \textbf{P3} ensures that \acron leaks nothing about \s besides its intended execution output, it fulfills condition (iii).
Also, \textbf{P4} guarantees that \s cannot be modified except by a fresh instance of \acron protocol, which satisfies condition (iv).
Meeting all conditions in Definition~\ref{def:sp-pox}, \acron is also SP-\pox-secure.
Lastly, \posx security (per Definition~\ref{def:posx}) follows directly from simultaneous adherence to IV-\pox and SP-\pox security.\qed

\textit{\textbf{Remark.} \acron maintains \textbf{P1}-\textbf{P6} even in the presence of an adaptive \adv; this implies that irrespective of \adv actions or knowledge of \acron (assuming no invasive hardware attacks), none of these properties can be compromised.}

\section{\acron Extensions and Variations}\label{sec:ext}


\textbf{Cryptographic Choices.}
Although Protocol~\ref{prot:posx} uses public-key cryptography,
it can seamlessly transition to symmetric cryptography by simply substituting public-key operations (i.e., \textsf{Sign} and \textsf{Verify}) with MAC operations.
If quantum threats are in scope, \acron can be similarly adjusted to support a post-quantum signature scheme.
We demonstrate this versatility by implementing our prototype (Section~\ref{sec:impl}) using three distinct cryptographic choices (public-key, symmetric, and post-quantum cryptography).

\textbf{Space-Time Trade-Off.}
Section~\ref{sec:posx_prot} discussed choices for managing $\s_{sec}$ in the Secure World: (1) storing the entire \s value; or (2) maintaining a hash of \s value.
The first prioritizes runtime efficiency since it requires no run-time hash computation while the second conserves storage in the Secure World by condensing the potentially large \s into a fixed-size digest. 
A third option, that eliminates storage in the Secure World, is to have \texttt{SetState} compute a MAC of \s (instead of a hash) and pass it back to the Non-Secure World for storage. Subsequently, the Secure World can authenticate \s received from the Normal World based on the MAC, yielding equivalent security guarantees as the previous two approaches.
By default, \acron prototype adopts the second design choice, striking a balance between space and time overhead.

\textbf{Multiple Stateful Functions.}
Our description of \acron assumes that \vrf intends to obtain one \posx per \prv at a time.
Nonetheless, it can be extended to accommodate simultaneous \posx-s by maintaining a map between multiple $\s_{sec}$-s and each ongoing \posx, in the Secure World.
To that end, \texttt{CheckState} and \texttt{SetState} should be extended to assign $\s_{sec} = map[id]$ for a given function identifier $id$.
Also, in the last step of the protocol, \vrf must additionally inspect \f binary to ensure that it correctly calls \texttt{CheckState} and \texttt{SetState} with the correct function identifier.

\textbf{Relaxing Atomicity Requirement.}
\acron security mandates atomicity (uninterruptability) during \f execution.
This requirement may clash with real-time needs on \dev, potentially preventing time-sensitive tasks from completing while \acron is running.
Recent studies~\cite{asap,iscflat} propose techniques to relax this atomicity requirement in classic \pox and related schemes. These can also be adopted in \acron as follows:
rather than completely disabling interrupts, the Secure World ``locks" (i.e., making them read-only) $PMEM$, data currently in use by the \posx task, and the Interrupt Vector Table (IVT).
It also includes IVT in the snapshot $h$ before invoking \f. 
As a consequence, the \posx context is protected across interrupts.
Once \f completes, respective memory can be unlocked.

\section{From \posx to Poisoning Prevention}\label{sec:poison}


We now discuss how \acron can be leveraged to detect poisoning attacks in \texttt{LDP-DC} and \texttt{FL-DC}.

\subsection{Poisoning-free LDP.}\label{sec:pf-ldp}
\revision{\texttt{LDP-DC$^+$} relies on \acron to ensure the correct execution of the following steps:}
\begin{enumerate}
    \item \textbf{Setup.} \revision{Run \acron protocol to obtain a \posx of \texttt{Init-state$_B$($\phi$)}, which executes the function \texttt{Init-state} without any input using the state variable $B$ (for $PRR$ mapping).
   This execution initializes $B$ to an empty list on \dev. Abort if the protocol outputs $\bot$.}
    \item \textbf{Collect.} \revision{Run \acron protocol to obtain a \posx of \texttt{LDP-DC$_B$($\f$,$\inp$,$f$,$p$,$q$)} as specified in Algorithm~\ref{alg:ldp}.
    On state $B$, this execution performs a sensor reading $\f(\inp)$, perturbs the reading result using the LDP-based mechanism with the parameter values $f$, $p$, and $q$, and returns the noisy output \outp to \vrf.
    Repeat this step if \vrf wants to collect more readings.} 
\end{enumerate}

\acron in \textbf{Setup} step ensures that the state variable $B$ is initialized to an empty value.
At a later time, \textbf{Collect} can be executed, where
\acron protocol gives assurance to \vrf that: (1) \outp is genuine, originating from a timely execution of the sensor function, and correctly privatized by the underlying LDP mechanism, (2) $B$ corresponded to the authentic value (e.g., empty at the first time of this step's execution) right before and during the protocol execution, and (3) $B$ is correctly updated as a result of the protocol execution.
These prevent poisoning attacks because \adv can tamper with neither \outp (during \textbf{Collect}) nor $B$ (during \textbf{Setup} or \textbf{Collect}).

\subsection{Poisoning-free FL.}\label{sec:pf-fl}
\texttt{FL-DC$^+$} follows a similar approach by using \acron to convey to \vrf the correct execution of:

\begin{enumerate}
    \item \textbf{Setup.} \revision{Run \acron protocol to obtain a \posx of \texttt{Init-State$_D$($\phi$)}, which executes without input arguments and uses the local training dataset $D$ as the underlying \posx state.
    The execution sets $D$ to an empty list on \dev. Abort if it outputs $\bot$.}
    \item \textbf{Collect.} \revision{To perform a sensor reading $\f(\inp)$ and record the authentic result to the state $D$, run \acron to obtain a \posx of \texttt{Sense-Store$_D$($\f$,$\inp$)}; see Algorithm~\ref{alg:fl}. Repeat this step to collect more sensor readings. Abort if the protocol outputs $\bot$.}
    \item \textbf{Train.} \revision{Run \acron protocol to obtain a \posx of \texttt{Train$_D$($W$,$t$,$\alpha$)} as specified in Algorithm~\ref{alg:fl}.
    This execution performs local training on the state $D$ using the \vrf-requested global weights $W$ and training parameters $t$ and $\alpha$; it then outputs the locally trained model \outp to \vrf.}
\end{enumerate}

Similar to \texttt{LDP-DC$^+$}, \textbf{Setup} guarantees to \vrf that $D$ starts empty. For \textbf{Collect}, \acron ensures that each record in $D$ is produced as a result of executing the expected sensor function on \dev. 
Finally, \textbf{Train} assures \vrf that the correct training function was applied to the authentic training data in $D$, leading to the received trained model \outp.
In the context of FL, \adv may attempt poisoning attacks in two ways: (1) model poisoning by directly tampering with the trained model \outp and (2) data poisoning by manipulating the dataset $D$ used for training. \revision{Both are prevented by \texttt{FL-DC$^+$}}.

\textit{\textbf{Remark.} We emphasize that both \texttt{LDP-DC$^+$} and \texttt{FL-DC$^+$} maintain the privacy of their original counterparts. The only additional information besides \outp obtained by \vrf is $\sigma$, which is not a function of the underlying execution state. Also, as a system-level approach, these schemes offer a deterministic guarantee in discerning poisoned data from authentic data, i.e., achieving 100\% true positive and true negative rates, irrespective of adaptive attacks. It also comes without any assumptions about the data distributions on \prv.}

\section{Evaluation}\label{sec:impl}

\subsection{Experimental Setup}

\textbf{Cryptographic Variants.}
As noted in Section~\ref{sec:ext}, \acron is agnostic to underlying cryptographic primitives.
To showcase this flexibility, we provide 3 implementation variants of \acron's RoT: $\acron_{SK}$, $\acron_{PK}$, and $\acron_{PQ}$:
\begin{enumerate}
    \item $\acron_{SK}$ uses symmetric-key cryptography to implement \textsf{Sign} and \textsf{Verify} in Protocol~\ref{prot:posx} with an HMAC-SHA256 in line with prior \pox work~\cite{apex}.

    \item $\acron_{PK}$ relies on the public-key signature ECDSA NIST256p from micro-ECC library\footnote{https://github.com/kmackay/micro-ecc}, which is commonly used in embedded settings~\cite{silde2019comparative, mossinger2016towards}.

    \item $\acron_{PQ}$ implements \textsf{Sign} and \textsf{Verify} using the quantum-resistant public-key signature Sphincs+, a low-RAM version of Sphincs-sha2-128f\footnote{https://github.com/sphincs/low-ram-sphincsplus}.
    This choice is supported by prior research~\cite{niederhagen2022streaming, hulsing2016armed} showing feasibility of Sphincs+ on Cortex-M devices.
\end{enumerate}

\textbf{Prototype.}
We prototype \acron variants on a NUCLEO-L552ZE-Q~\cite{nucleo} development board, representing resource-constrained IoT devices.
It features TrustZone-M on an ARM Cortex-M33 MCU @ 110MHz clock, with 512KB of FLASH (of which we assign 256KB to store the Non-secure World's PMEM) and 256KB of RAM.
To accurately isolate \acron overheads, we implement a simple stateful \f function on \dev that takes input from a GPIO port specified from a \posx request, performs a sensor reading on that port, and outputs an accumulative sum of all readings over time to \vrf.
\vrf is deployed as a commodity desktop equipped with an Intel i5-9300H CPU @ 2.4GHz. 
\vrf and \dev are connected via serial communication. All prototypes are open-sourced and publicly available at~\cite{repo}.

\subsection{Baseline.}
For comparison, we consider an alternative naive baseline approach in which all functions to prove execution \f are included as part of TCB in the Secure World.
To perform \posx, TCB receives and authenticates a request from the Non-Secure World, just like \acron. 
Unlike \acron, this approach executes \f inside the Secure World, 
The result is signed (or MAC'ed) and forwarded to \vrf, akin to \acron.
As this baseline approach is also agnostic to the underlying cryptographic primitives,
we refer to $\sf{Baseline_{SK}}$, $\sf{Baseline_{PK}}$ and $\sf{Baseline_{PQ}}$ as the baseline approaches that utilize symmetric-key cryptography (HMAC-SHA256), public-key cryptography (ECDSA) and post-quantum cryptography (Sphincs+), respectively.

We note that this baseline faces several security and practical downsides. 
First, its TCB becomes bloated and dependent on multiple untrusted applications by including all \f-s within the Secure World.
This implies that a vulnerability in one of them can compromise all (i.e., violating the principle of least privilege).
It also incurs Secure World's additional storage for maintaining data of \f execution.
This reduces available RAM for normal applications in the Non-Secure World.
Moreover, it makes the Secure World code (which should be immutable post-deployment -- recall Section~\ref{sec:bg_tz}) application-specific and thus requires rewriting/updating the Secure World every time to support new applications, which may necessitate cumbersome physical intervention.

\begin{table}[]
\centering
\caption{Binary size (in KB) of TCB.}
\label{tab:code-size}
\resizebox{.7\columnwidth}{!}{%
\begin{tabular}{c|cc|c}
\hline
Variants       & $\sf{Baseline}$ & \acron & Overhead  \\ \hline\hline
Symmetric-key & 17.0                  & 17.5                 & 2.9\% \\
Publick-key & 34.5                  & 35.0                 & 1.4\%  \\
Post-quantum & 24.0                  & 24.5                 & 2.0\%   \\ \hline
\end{tabular}%
}
\vspace{-1em}
\end{table}

\subsection{Space Overhead}

\textbf{Code Size.}
\acron's TCB was implemented in C and compiled using the \texttt{-O3} optimization flag.
Details of code size are shown in Table~\ref{tab:code-size}. 
As $\acron_{SK}$ relies on an inexpensive cryptographic primitive, it exhibits the smallest code size.
Conversely, $\acron_{PK}$ results in the largest binary size of 35.0KB while $\acron_{PQ}$'s binary is around 24.5KB.
Compared to the baselines, \acron introduces a small code size overhead (0.5KB), corresponding to 1.4-2.9\% across all variants.

For \f instrumentation, \acron requires prepending a call to \texttt{CheckState} at the beginning of \f and another call to \texttt{SetState} before \f returns.
This instrumentation results in only 2 additional lines of C code, enlarging \f's binary (residing in Non-Secure World) by a fixed 22 bytes.

\begin{figure}
\begin{subfigure}[h]{0.46\linewidth}
\includegraphics[width=\linewidth]{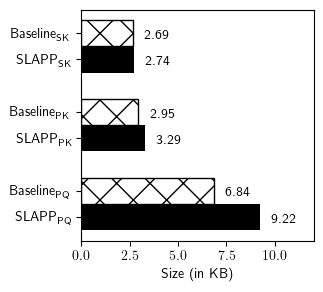}
\caption{Peak runtime data allocation}
\label{fig:ram}
\end{subfigure}
\hfill
\begin{subfigure}[h]{0.46\linewidth}
\includegraphics[width=\linewidth]{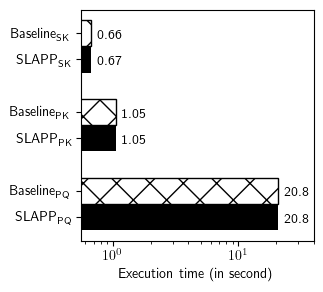}
\caption{Average execution time}
\label{fig:runtime}
\end{subfigure}%
\caption{Resource usage across all approaches}
\end{figure}

\ignore{
\begin{figure}[h]
    \centering
    \includegraphics[width=\linewidth]{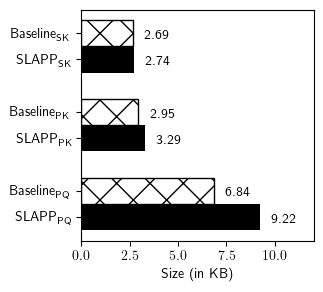}
    \caption{Peak runtime data allocation across all approaches}
    \label{fig:ram}
\end{figure}
}

\textbf{Memory Usage.}
Next, we estimate the peak amount of Secure World data allocated at runtime.
This data consists of all stack and static/global variables (our implementation utilizes no heap allocation).
As shown in Fig.~\ref{fig:ram}, $\acron_{SK}$ and $\acron_{PK}$ uses roughly the same amount of data: 2.74 and 3.29KB. $\acron_{PQ}$, on the other hand, requires more than triple this amount: 9.22KB.
Compared to the baselines, our \acron variants incur an additional memory usage ranging from 50 bytes for $\acron_{SK}$ to 2.38KB for $\acron_{PQ}$.
These correspond to $<$ 4\% of available RAM in the Secure World.

\ignore{
\begin{figure}[h]
    \centering
    \includegraphics[width=\linewidth]{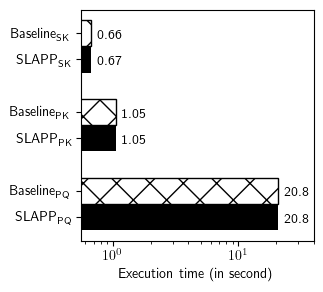}
    \caption{Execution time across all approaches}
    \vspace{-1em}
    \label{fig:runtime}
\end{figure}
}

\subsection{Time Overhead}
As \vrf operates on a powerful back-end, its runtime within the \acron protocol is negligible compared to the execution time on \dev.
Thus, we focus on measuring \dev's execution time (i.e., the runtime time of executing Phases 3, 4, and 5 in Section~\ref{sec:posx}).
Results are illustrated in Fig.~\ref{fig:runtime}.

As expected, $\acron_{SK}$ has the fastest runtime: 0.67$s$. 
In contrast, $\acron_{PQ}$ utilizes expensive cryptography and thus incurs the longest execution time of $\approx20s$.
$\acron_{PK}$ positions between these two, taking around 1 second.

\begin{figure}[h]
    \centering
    \includegraphics[width=.9\linewidth]{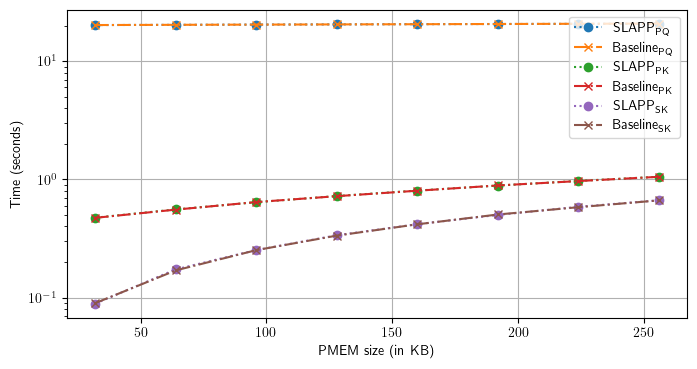}
    \caption{Execution time with varying $PMEM$ size}
    \label{fig:rtpmem}
\end{figure}

Since \acron requires \dev to compute a snapshot of $PMEM$ as part of proof generation, we conducted an experiment to assess the impact of $PMEM$ size on \dev's execution time.
In this experiment, we varied the $PMEM$ size from 32KB to 256KB.
As depicted in Fig.~\ref{fig:rtpmem}, the time to execute this snapshot contributes significantly to the overall execution time when $\acron_{SK}$ is used.
It also shows a linear relationship between execution time and $PMEM$ size for $\acron_{SK}$ and $\acron_{PK}$.
However, for $\acron_{PQ}$, this effect is negligible; its runtime has minimal impact on the overall execution time regardless of $PMEM$ size.

From Figs.~\ref{fig:runtime} and~\ref{fig:rtpmem}, we can conclude that all \acron variants incur a negligible execution time overhead, i.e., $<1\%$, compared to their baseline counterparts. We next conduct end-to-end evaluation of \acron through two case studies.

\subsection{Case Study 1: Local Differential Privacy}

\textbf{Description.}
In the first case study, we envision the integration of \acron within a smart city/grid system. 
The service provider aims to periodically collect energy consumption data from all smart meters located in individual households to calculate electric bills, forecast load, etc. 
Previous studies~\cite{mcdaniel2009security, molina2010private} have demonstrated privacy risks by exposing raw energy to the service provider, e.g., with access to such data, the provider could potentially infer users' habits and behavior.
\texttt{LDP-DC} (from Algorithm~\ref{alg:ldp}) can be employed to address this concern. 
The service provider is also motivated to use a poisoning-free version, \texttt{LDP-DC$^+$} as presented in Section~\ref{sec:pf-ldp}, to prevent poisoning attacks from potentially malicious edge devices.

\begin{table*}[]
\centering
\caption{End-to-end results for $\sf{Baseline_{SK}}$ (shown in parentheses) and $\acron_{SK}$. Phases 1-5 are as defined in Section~\ref{sec:posx_prot}.}
\label{tab:case-study}
\vspace{-2mm}
\resizebox{\linewidth}{!}{%
\begin{tabular}{l|c|c|cccccc|cc|cc}
\hline
\multicolumn{1}{c|}{\multirow{2}{*}{Case Study}} &
  \multirow{2}{*}{Protocol} &
  \multirow{2}{*}{Step} &
  \multicolumn{6}{c|}{Average Execution Time (ms)} &
  \multicolumn{2}{c|}{RAM (KB)} &
  \multicolumn{2}{c}{FLASH (KB)} \\
\multicolumn{1}{c|}{} &  &                  & Phase 1 & Phase 2 & Phase 3 & Phase 4 & Phase 5 & Total & S & NS & S & NS \\ \hline\hline
\multirow{2}{*}{\begin{tabular}[c]{@{}l@{}}(1) Collect noisy\\ energy readings\end{tabular}} &
  \multirow{2}{*}{\texttt{LDP-DC$^+$}} &
  \textbf{Setup} & 8 & 660   & 7   & 1 & 11  & 687(678)   &  4.7(4.9)   & 2.7(1.2)   & \multirow{2}{*}{17.4(23.2)}   & \multirow{2}{*}{8.5(7.4)}  \\
                      &  & \textbf{Collect} &  6       &    660     &     9    &    1     &  8       &   684(677)    & 4.7(4.9)     &  2.7(1.2)        &      &          \\ \hline
\multirow{3}{*}{\begin{tabular}[c]{@{}l@{}}(2) Train a model\\ for load forecasting\end{tabular}} &
  \multirow{3}{*}{\texttt{FL-DC$^+$}} &
  \textbf{Setup} & 13 & 652   & 8   & 1   & 8   & 682(678)  & 4.7(5.4)  & 3.3(1.5)  & \multirow{3}{*}{17.4(71.3)}   & \multirow{3}{*}{57.2(7.5)}    \\
                      &  & \textbf{Collect} &  12       &   654      &   8      &   1      &   7      &   682(682)   &  4.7(5.4)   &  3.3(1.5)        &      &          \\
                      &  & \textbf{Train}   &  16       &  651       &  3,067       &   2      &    20     &  3,756(3,759)     &  4.7(32.0)    &   32.0(1.5)       &      &          \\ \hline
\end{tabular}%
}
\vspace{-2mm}
\end{table*}

We build a prototype of a smart meter (\dev) based on NUCLEO-L552ZE-Q connecting to a PZEM-004T energy meter hardware module.
\dev's Non-Secure World consists of \texttt{LDP-DC} software (Algorithm~\ref{alg:ldp}) and a driver responsible for retrieving energy data from the PZEM-004T hardware module.
Meanwhile, the Secure World contains $\acron_{SK}$'s TCB.
The service provider (\vrf) runs on a commodity desktop and communicates with \dev over serial communication.

During normal (benign) operation, \vrf and \dev execute an instance of \texttt{LDP-DC$^+$} protocol.
This instance begins with the \textbf{Setup} step, which requires \posx of \texttt{Init-State} function to initialize a \dev-local state ($PRR$ mapping) to zeroes.
\vrf verified that \dev faithfully executes this step, completing with $\top$.
Upon obtaining the successful output, \vrf proceeds to the \textbf{Collect}, which aims to collect authentic noisy energy data from \dev.
To achieve this, \vrf makes a \posx request by configuring a function to prove execution to \texttt{LDP-DC} and properly selecting LDP input parameters (i.e., $f$, $p$ and $q$) to meet the $\epsilon$-LDP requirement, and sends an authenticated \posx request to \dev. 
To successfully respond to this \posx request, \dev must execute \texttt{LDP-DC} correctly (with \vrf-defined inputs) and return the result (authentic noisy energy data) to \vrf.
If this step completes with $\top$, it ensures \vrf that the received energy data is not poisoned.

\textbf{End-to-end Evaluation.}
Results for this case study are presented in Table~\ref{tab:case-study}. 
$\acron_{SK}$ takes around 17.4KB and 8.5KB of Secure and Non-Secure FLASH, respectively. 
$\sf{Baseline_{SK}}$ would require placing the code implementing the LDP logics into Secure FLASH, enlarging its TCB by 5.8KB or 33.3\%. Note that this number would further increase with complexity of functionality or if multiple different sensing functions need to be implemented by \prv.
This result substantiates \acron's design rationale: being application-agnostic significantly reduces the TCB size and provides flexibility. 
Regarding RAM usage, $\acron_{SK}$ requires 4.7KB of RAM to execute the \textbf{Setup} and \textbf{Collect} steps, which is 0.2KB less than $\sf{Baseline_{SK}}$, leading to 4.3\% reduction in Secure RAM usage.


For both steps, the end-to-end execution time (from \vrf generating a request to \vrf verifying the response) is approximately 0.69 seconds, corresponding to 1\% increase over $\sf{Baseline_{SK}}$.
We also reported the execution time breakdown for all phases in Table~\ref{tab:case-study}. Recall Section~\ref{sec:posx_prot} for the definition of each phase.
As Phase 2 requires a hash computation over the 
Non-Secure FLASH of 256KB, 
it dominates the end-to-end runtime, accounting for $\approx$96\% of the overall runtime.
Phase 4 is the fastest, as it consists of only lightweight operations, i.e., setting flags and computing HMAC on a short message, without requiring any cross-world switching or network communication. 
Finally, the time taken by \vrf (Phases 1 and 5) contributes with $<$ 3\% of the overall runtime.

\begin{figure}
\begin{subfigure}[h]{0.475\linewidth}
\includegraphics[width=\linewidth]{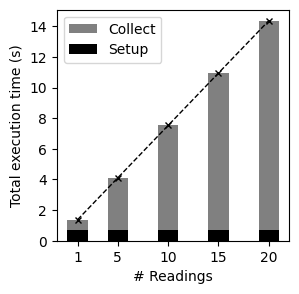}
\caption{Case study 1: \texttt{LDP-DC$^+$}}
\label{fig:ldp-runtime}
\end{subfigure}
\hfill
\begin{subfigure}[h]{0.45\linewidth}
\includegraphics[width=\linewidth]{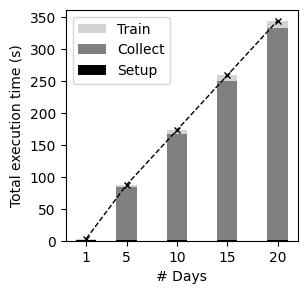}
\caption{Case study 2: \texttt{FL-DC$^+$}}
\label{fig:fl-runtime}
\end{subfigure}%
\caption{Sum of the runtimes of multiple data collection rounds performed over a longer period with multiple readings. 
}
\end{figure}

We next consider a task of using \texttt{LDP-DC$^+$} to collect a variable number of noisy energy readings from 1 to 20. This emulates the case of continuous data collection to be performed over a longer period of time. For instance, \vrf requests 1 reading every 3 minutes, resulting in 20 readings over an hour.
The results are shown in Fig.~\ref{fig:ldp-runtime}.
As this task requires invoking \textbf{Setup} once before \textbf{Collect} can be repeated as many times as needed, the time for \textbf{Setup} remains constant, irrespective of the number of subsequent collections. 
Conversely, the runtime overhead for completing all \textbf{Collect} steps increases linearly with the number of readings collected. We do not observe a significant runtime overhead of $\acron_{SK}$ compared to $\sf{Baseline_{SK}}$ in this case.

\textbf{Attack Simulation.}
We launch the following attacks to \texttt{LDP-DC$^+$} in this case study: 

    \noindent $\bullet$ $\adv_1$ corrupts \texttt{Init-State} to poison the initial state.
    
    \noindent $\bullet$ $\adv_2$ corrupts \texttt{LDP-DC} between \textbf{Setup} and \textbf{Collect} phases to poison raw energy readings.
    
    \noindent $\bullet$ $\adv_3$ poisons the state in between \textbf{Setup} and \textbf{Collect} phases.
    
    \noindent $\bullet$ $\adv_4$ poisons the final result in \textbf{Collect} phase.

\texttt{LDP-DC$^+$} detects all aforementioned attacks. 
Specifically, $\adv_1$ and $\adv_2$ are detected at the end of \textbf{Setup} and \textbf{Collect} phases, respectively.
$\adv_3$ is caught in \textbf{Collect} since \acron's RoT noticed tampered state values (from \texttt{CheckState}) and thus refused to generate a valid proof.
Similarly, $\adv_4$ is detected in \textbf{Collect} phase since the proof $\sigma$ does not reflect the tampered output received by \vrf.



\subsection{Case Study 2: Federated Learning}

\textbf{Description.}
\revision{
Here \vrf aims to develop an ML model for one-hour-ahead load forecasting~\cite{taik2020electrical}.
To achieve this while ensuring user privacy, it employs FL based on LSTM~\cite{taik2020electrical}. 
As a simple embedded device, \dev is not designed to handle large or complex ML models due to constraints on CPU, memory, and energy.
For example, the prototype board used in this case study operates at a 110MHz CPU clock speed and has only 256KB of RAM and 512KB of FLASH shared between the Secure and Non-secure Worlds.
These limitations also prohibit the storage and processing of a large training dataset.
To overcome these challenges, this case study restricts each \dev to collecting 5 days worth of hourly energy readings (i.e., 120 data points) and training a lightweight LSTM model with a single layer of 8 neurons. Once local training is complete, the local models are transmitted to and aggregated by \vrf.
\texttt{FL-DC$^+$} is employed to prevent poisoning.}


\dev and \vrf initiate \texttt{FL-DC$^+$} protocol by running \textbf{Setup} step, which clears all energy readings on \dev.
Next, \vrf periodically requests \textbf{Collect} to record an energy reading into a \dev-local training dataset.
Once \dev records enough energy readings, \vrf triggers the \textbf{Train} phase by transmitting a \posx request to \dev. This request specifies the initial weights to be trained on, the learning rate (0.01), and the number of epochs (5). \dev then (provably) performs the local training. 
Upon completion, \dev sends back the updated weights along with the proof of training to \vrf.
Since \vrf and \dev adhere to \texttt{FL-DC$^+$}, the protocol yields $\top$, ensuring to \vrf that the received model was correctly trained on authentic energy data and thus can be securely aggregated onto the global model.

\textbf{End-to-end Evaluation.}
The results of this case study are shown in Table~\ref{tab:case-study}. Similar to the previous case study, by making the TCB application-agnostic, $\acron_{SK}$ can reduce the size of Secure FLASH by a significant amount (76\%). $\acron_{SK}$ requires the same amount of Secure RAM regardless of the steps. In contrast, $\sf{Baseline_{SK}}$ incurs 0.7KB of Secure RAM for the \textbf{Setup} and \textbf{Collect} steps, while the \textbf{Train} step, which involves training an LSTM model, requires a more substantial amount, 32KB, of Secure RAM -- around 5\texttt{x} of $\acron_{SK}$. These results further emphasize \acron's greater benefits especially when \f is more resource-intensive, as in the case of LSTM training.

In terms of execution time, the results for the \textbf{Setup} and \textbf{Collect} steps are similar to the previous case study, with Phase 2 being the most time-consuming. However, the end-to-end execution time for \textbf{Train} is dominated by Phase 3, which involves executing \f (LSTM training) in the Non-Secure World.
This step takes around 3 seconds, contributing $\approx$81\% to the end-to-end runtime.

Finally, we consider a task of applying \texttt{FL-DC$^+$} to collaboratively train an LSTM model on hourly energy readings collected over different numbers of days (from 1 to 20 days).
The sums of total runtimes for the entire periods are shown in Fig.~\ref{fig:fl-runtime}. 
Since \textbf{Setup} is performed once, its overall runtime is fixed to 0.69s.
As the energy sensor is read 24 times a day, the time sum for all \textbf{Collect} steps is linear with the number of days and dominates the overall runtime. 
The \textbf{Train} step is executed at the end of this task on all previously collected readings, resulting in a linear runtime sum ranging from $2s$ (over 2 days) to $12s$ (over 20 days).
With most steps exhibiting a linear runtime, the total runtime of this task also becomes linear with the number of days.

\begin{figure}[h]
    \centering
    \includegraphics[width=.9\linewidth]{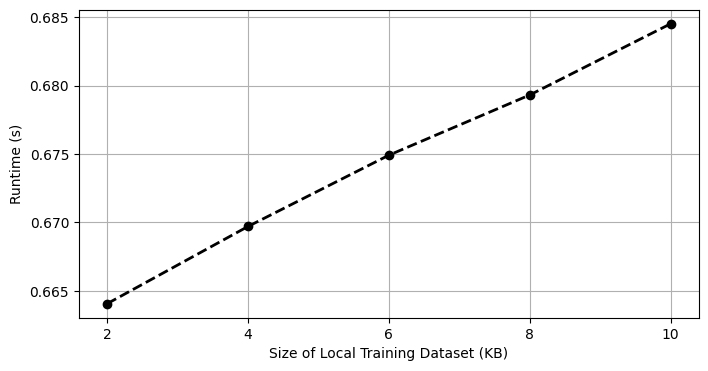}
    \caption{\dev runtime overhead with varying local training sizes}
    \label{fig:fl-runtime-statesize}
\end{figure}

\revision{
\textbf{Runtime Overhead vs Training Size.}
Next, we analyze the impact of the training dataset size on the runtime overhead introduced by \texttt{FL-DC$^+$}.
On \dev, compared to vanilla FedAVG, \texttt{FL-DC$^+$} added overhead corresponds to context switches between Secure and Non-Secure Worlds plus time to hash $PMEM$, authenticate \vrf request, compute one MAC/signature, and execute one \texttt{checkState}/\texttt{setState} (one hash computation for each). 
Among these, only the last operation depends on the size of the training dataset, which is the \posx state in \texttt{FL-DC$^+$}.
In this experiment, we consider larger (local) dataset sizes ranging from 2KB to 10KB. Assuming one sensor reading is collected per hour, these datasets correspond to 2.8-14 months of sensor data collection in this case study.
The experimental results, shown in Fig.~\ref{fig:fl-runtime-statesize}, demonstrate that \acron runtime overhead is linear in terms of the dataset size.
Compared to the local training time of around 23$s$ throughout this experiment, the added overhead ($\approx~0.6s$) is small: around $2.6\%$ of this duration.
We also note that since \vrf operates independently of the training dataset, no additional runtime is incurred on \vrf.
}

\begin{figure}[h]
    \centering
    \includegraphics[width=.9\linewidth]{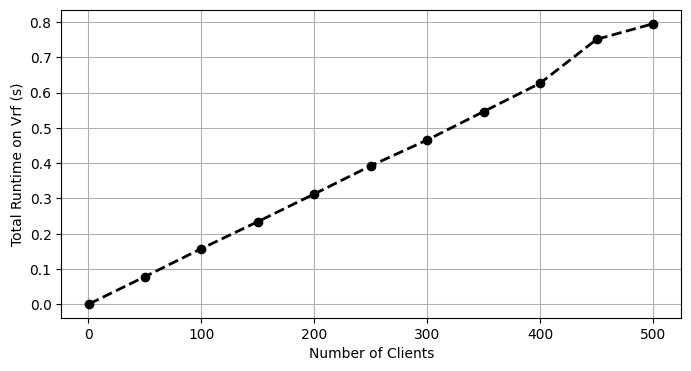}
    \caption{Total runtime on \vrf with multiple clients in \texttt{FL-DC$^+$}}
    \label{fig:fl-vrf-runtime}
\end{figure}

\revision{
\textbf{Scalability.} 
To evaluate the scalability of \texttt{FL-DC$^+$}, we measure the total runtime required for \vrf to execute this case study with a variable numbers of clients (\dev-s). 
Upon receiving \posx-s of the \textbf{Train} phase from all clients, \vrf performs one authentication for each \posx and then aggregates all local models that pass the authentication checks into the current global model.
We report the total runtime results on \vrf in Fig.~\ref{fig:fl-vrf-runtime}.
Since the runtime linearly depends on the number of \posx-s received, which equals the number of clients, it scales linearly with the number of clients. Notably, even for 500 clients, \vrf completes its operations in less than a second.
}

\begin{figure}[h]
    \centering
    \includegraphics[width=.9\linewidth]{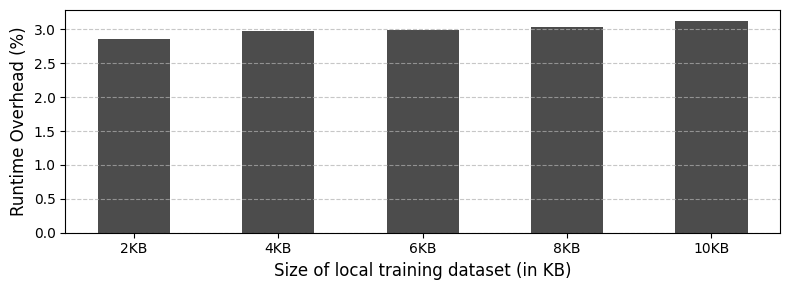}
    \caption{\texttt{FL-DC$^+$} worst-case runtime overhead w.r.t. existing poisoning prevention techniques in FL}
    \label{fig:fl-vrf-runtime-existing}
\end{figure}

\revision{
\textbf{Comparison with Existing Techniques.}
Here, we compare the runtime of \texttt{FL-DC$^+$} with existing techniques for mitigating poisoning attacks in FL.
We consider two types of existing techniques: the ones with modified loss functions~\cite{feng2014robust, jagielski2018manipulating,sun2021fl} and the ones based on Byzantine-robust aggregation~\cite{blanchard2017machine,guerraoui2018hidden}.
In the worst case, \texttt{FL-DC$^+$} overhead is incurred due to three additional computations on \dev: one for hashing $PMEM$, one hash during \texttt{CheckState} and another during \texttt{SetState}\footnote{Note that \vrf authentication and the time to produce a MAC/signature are not considered \texttt{FL-DC$^+$} overhead since these operations are also required by existing techniques, e.g., the Byzantine-robust techniques need mutual authentication to prevent a malicious \dev from impersonating others and breaking the Byzantine threshold.}.
Thus, the worst-case runtime overhead of \texttt{FL-DC$^+$} also depends on the size of the training dataset.
To evaluate this, we measured the overhead for training datasets ranging from 2KB to 10KB.
As seen in Fig.~\ref{fig:fl-vrf-runtime-existing},
our approach introduces around 3\% runtime overhead.

In terms of storage overhead, \acron additionally requires its (application-agnostic) TCB to be implemented inside TrustZone-M Secure World. With the symmetric-key version ($\acron_{SK}$), this overhead results in 17.5KB of additional FLASH secure storage.
Given its small overheads, we emphasize that \texttt{FL-DC$^+$} need not (and is not meant to) replace prior techniques and can be combined with them for increased security; we elaborate on this point in Section~\ref{sec:acron-with-others}.
}

\textbf{Attack Simulation.}
We simulate two types of adversaries: $\adv_D$ and $\adv_M$. $\adv_D$ performs data poisoning attacks to \texttt{FL-DC$^+$} by compromising \dev-local training dataset before \textbf{Train} is executed.
Meanwhile, $\adv_M$ simulates model poisoning by manipulating the locally trained model parameters before the model arrives at \vrf.
\texttt{FL-DC$^+$} detects $\adv_D$ during the \textbf{Train} phase because \acron's RoT identifies state \s tampering (\s is the training dataset in this case). $\adv_M$ is caught at the end of the \textbf{Train} as the proof does not match the tampered output.

\revision{
\section{Integrating Complimentary Techniques with \acron for Additional Benefits}\label{sec:acron-with-others}

As mentioned in Section~\ref{sec:intro}, \acron offers two notable advantages: (1) it allows \dev to convince \vrf that execution of a function \f happened without any input/state assumption about \f and (2) most \acron operations are performed on the client-side (i.e., \dev), making it possible to combine \acron with any server-side techniques.
In this section, we present two concrete examples that leverage these advantages to enhance \acron benefits beyond poisoning protection in the scope of our system model. 
We note that while we focus on \texttt{FL-DC$^+$}, the discussion in this section is also relevant to \texttt{LDP-DC$^+$}.

First, \texttt{FL-DC$^+$} builds upon FedAVG algorithm, which is shown to perform poorly with non-identically distributed (IID) data across \dev-s~\cite{zhao2018federated}. 
To cope with this, several studies have extended FedAVG to better handle non-IID data.
For instance, FedProx~\cite{fedprox} introduces a proximal term to the loss function during the local training process to help constrain local updates to be closer to the global model.
As \acron supports \posx for arbitrary \f, 
\texttt{FL-DC$^+$} can be adapted to incorporate FedProx by implementing the FedProx algorithm as \f in the \textbf{Train} phase.
With this minor modification, \texttt{FL-DC$^+$} extends poisoning prevention to the non-IID setting.

Second, \acron threat model focuses on software-only attacks while considering attacks that manipulate \dev hardware or the physical environment being measured by \dev (e.g. the ones considered in~\cite{gu2019badnets,wenger2021backdoor}) out of scope.
As \texttt{FL-DC$^+$} builds atop \acron, it inherits the same assumption for poisoning prevention.
One common approach to address poisoning attacks in FL against hardware/physical attacks is through Byzantine-robust aggregation techniques~\cite{blanchard2017machine,guerraoui2018hidden}.
These techniques modify the aggregation step on \vrf to make it more robust against malicious updates under the Byzantine assumptions (i.e., only a certain number of \dev-s can be compromised at a time), alleviating the impact from (but not completely preventing) hardware/physical \adv.
For example, the work in~\cite{chen2017distributed} replaces the arithmetic mean of local gradients (as used in FedAVG) with the geometric median of means during global model updates.
If hardware/physical attacks are of concern (in addition to software-based attacks), \texttt{FL-DC$^+$} can be adapted to support Byzantine robustness mechanisms.
In particular, after receiving all verified local models from \dev-s, \vrf in \texttt{FL-DC$^+$} can update the global model via Byzantine-robust aggregation rules.
This helps mitigates poisoning attacks from Byzantine hardware/physical \adv in addition to software-only \adv. We believe the combination of both approaches to significantly strengthen overall security of these schemes.
}



\section{Related Work}





\begin{table*}[]
\centering
\caption{Comparison with current defenses (\emptycirc, \halfcirc~and \fullcirc~indicates the degree of support/assumption/impact)}
\label{tab:comparison}
\resizebox{\textwidth}{!}{%
\begin{tabular}{lc|cc|cc|cc|cc}
\hline
\multicolumn{2}{c|}{\multirow{3}{*}{\normalsize \shortstack[c]{Feature ($\rightarrow$) \\ Approach ($\downarrow$)}}} &

  \multicolumn{2}{c|}{\normalsize \adv} &
  \multicolumn{2}{c|}{\normalsize Data} &
  \multicolumn{2}{c|}{\normalsize Application} &
  \multicolumn{2}{c}{\normalsize \dev} \\
\multicolumn{2}{l|}{} &
  \begin{tabular}[c]{@{}c@{}}Adaptive\\ Robustness\end{tabular} &
  \multicolumn{1}{c|}{\begin{tabular}[c]{@{}c@{}}Byzantine\\Assumption\end{tabular}} &
  \begin{tabular}[c]{@{}c@{}}Distribution\\ Assumption\end{tabular} &
  \multicolumn{1}{c|}{\begin{tabular}[c]{@{}c@{}}Impact on\\ Agg. Utility\end{tabular}} &
  \begin{tabular}[c]{@{}c@{}}Support\\ DP/LDP\end{tabular} &
  \begin{tabular}[c]{@{}c@{}}Support\\ FL/ML\end{tabular} &
  \begin{tabular}[c]{@{}c@{}}Require\\ RoT\end{tabular} &
  \begin{tabular}[c]{@{}c@{}}App-indep.\\ RoT\end{tabular} 
   \\ \hline \hline
\multirow{4}{*}{Data-driven}  & Sniper~\cite{cao2019understanding}   & \emptycirc  & \fullcirc & \fullcirc & \fullcirc & \emptycirc & \fullcirc & \emptycirc & N/A  \\ 
                              & Normalize.~\cite{cao2021data}      & \halfcirc   & \fullcirc & \fullcirc & \fullcirc & \fullcirc & \emptycirc & \emptycirc & N/A \\ 
                              & ERR+LFR~\cite{fang2020local}       & \emptycirc & \fullcirc & \fullcirc & \fullcirc & \emptycirc & \fullcirc & \emptycirc & N/A \\ 
                              & Clustering~\cite{li2023fine}           & \emptycirc & \fullcirc & \fullcirc & \fullcirc & \fullcirc & \emptycirc & \emptycirc & N/A \\\hline 
                              
\multirow{4}{*}{Algorithmic}  & Multi-Krum~\cite{blanchard2017machine}  & \emptycirc  & \fullcirc & \fullcirc & \fullcirc & \emptycirc & \fullcirc & \emptycirc & N/A   \\ 
                              & RoLR~\cite{feng2014robust}     & \emptycirc   & \fullcirc & \fullcirc & \fullcirc & \emptycirc & \fullcirc &  \emptycirc & N/A   \\ 
                              & TRIM~\cite{jagielski2018manipulating} & \emptycirc  & \fullcirc & \emptycirc & \fullcirc & \emptycirc & \fullcirc &  \emptycirc & N/A   \\ 
                              & FL-WBC~\cite{sun2021fl}      & \halfcirc   & \emptycirc & \emptycirc & \fullcirc & \emptycirc & \fullcirc &  \halfcirc & \emptycirc  \\\hline 
                              
\multirow{2}{*}{System-level} & CrowdGuard~\cite{rieger2022crowdguard} & \halfcirc & \fullcirc & \emptycirc & \fullcirc & \emptycirc & \fullcirc &  \fullcirc & \emptycirc   \\ 
                              & \revision{EBFA~\cite{ebfa}}                       & \emptycirc & \fullcirc & \emptycirc & \fullcirc & \emptycirc & \fullcirc &  \fullcirc & \emptycirc   \\
                              & \acron (this work)                       & \fullcirc    & \emptycirc & \emptycirc & \emptycirc & \fullcirc & \fullcirc &  \fullcirc & \fullcirc  \\\hline 
\end{tabular}%
}
\end{table*}

{\bf Poisoning Prevention in LDP and FL.} 
We divide existing approaches to preventing poisoning attacks into 3 categories: data-driven, algorithmic and system-level.
Data-driven approaches detect poisoning attacks based solely on the collected data without modifying the underlying collection scheme, e.g., by applying normalization~\cite{cao2021data} and analyzing distances between or error rates from data~\cite{cao2019understanding, li2023fine,fang2020local}.
Meanwhile, algorithmic approaches modify the data collection algorithm to make it resilient against poisoning attacks, e.g., by incorporating with Byzantine fault-tolerant techniques~\cite{blanchard2017machine,guerraoui2018hidden,ebfa} or using modified loss functions~\cite{feng2014robust, jagielski2018manipulating,sun2021fl}.
Finally, system-level approaches~\cite{rieger2022crowdguard} leverage \dev's security architecture to mitigate poisoning attacks. \acron falls into the last category.

Based on these categories, we qualitatively compare \acron with current defenses in Table~\ref{tab:comparison}.
It shows that \acron is the only approach that offers strong robustness against adaptive adversaries without making assumptions about Byzantine \adv (i.e., \acron allows \adv to corrupt any number of \dev-s) or data distributions (i.e., \acron allows each \prv to have independent data distributions).
As \acron can deterministically discern poisoned data from the benign, it results in no utility loss on aggregate data.
It also supports both LDP and FL-based data collection schemes.
As a system-level approach, \acron requires \prv to have an RoT, which is becoming more common in modern IoT devices.
Moreover, \acron's RoT hosts only a small TCB agnostic to the data collection scheme (FL or LDP); this enables a \emph{one-and-done} process for validating its correctness 
(e.g., through formal verification, which is an interesting avenue for future work).
Besides, to the best of our knowledge, none of existing work provides \posx-equivalent guarantee of provable integrity all the way from data acquisition until its {\it de facto} usage as an aggregated statistical result or global ML model. They also do not address the issues of input validation or state preservation for arbitrary functions \f to be executed on simple \prv devices.

{\bf Verifiable Software Integrity in Embedded Devices.}
To secure low-end embedded devices, various 
low-cost security architectures have been proposed for remote verification of their software state via integrity proofs~\cite{mini-survey}. These proofs vary in terms of expressiveness, with simpler ones confirming 
correct binary presence (remote attestation)~\cite{vrased, smart, tytan, trustlite, hydra, Sancus17}, while more expressive ones support verification of arbitrary code execution. Aside from \pox architectures ~\cite{apex,asap,rares,flicker}, control flow attestation/auditing (CFA) techniques~~\cite{iscflat, cflat, scarr, recfa, oat, wang2023ari,lofat, dessouky2018litehax, zeitouni2017atrium,tinycfa,acfa} prove to \vrf the exact order in which instructions have executed within a particular code in \prv, thus enabling detection of code reuse attacks that can be triggered if the code whose execution is being proven is itself vulnerable. Data flow attestation~\cite{oat, dialed, dessouky2018litehax} augments CFA to generate evidence about memory safety violations even when exploits do not alter a program's legal control flow path.

{\bf Private Data Collection on Edge Devices.} Complementary to privacy mechanisms focusing on hiding private data from back-ends (e.g., LDP/FL), recent work has delved into assuring that private data is secure against compromised sensing devices ``from its birth'', i.e., from the moment when it is digitized. VERSA \cite{versa} was proposed as a HW/SW architecture to guarantee that only the correct execution of expected and explicitly authorized software can access and manipulate sensing interfaces. As a consequence, it blocks malware/modified software from accessing sensitive sensed quantities by default. Following this notion, Sensing And Actuation As A Privilege (SA4P)~\cite{SA4P,de2024sa} realizes this concept using ARM TrustZone.

\revision{\textbf{System-level Approaches in FL.}
Besides poisoning prevention, several system-level approaches have been proposed to provide different security and privacy guarantees in FL for higher-end \dev-s.
In terms of privacy, PPFL~\cite{mo2021ppfl} introduces a layer-wise training method within a Trusted Execution Environment (TEE) to provide confidentiality of training data while adhering to the memory constraints of the TEE. GradSec~\cite{messaoud2022shielding} improves upon this approach by significantly reducing the runtime overhead associated with the training process. 
Also, other approaches such as EBFA~\cite{ebfa}, CrowdGuard~\cite{rieger2022crowdguard} and Hashemi et al.~\cite{hashemi2021byzantine} propose the use of TEE to protect privacy of the training data/model while running poisoning prevention techniques outside the TEE.
For security, Pelta~\cite{queyrut2023mitigating} leverages TEE to mitigate adversarial (a.k.a. evasion) attacks in FL. It securely hides critical model parameters and updates these parameters (i.e., backpropagration) inside the client-side TEE. As a result, it limits a malicious client's access to only a partial model, making it harder to craft adversarial examples.
While sharing the similarity of leveraging TEE, these approaches do not use TEE to address data/model poisoning attacks in FL, which is one of the focal points in this work.
}


\section{Conclusion}

We defined and developed stateful proofs of execution, a system security primitive to thwart poisoning in applications such as differential privacy and federated learning. We analyze the security of our design (\acron) and evaluate its performance with an open-source prototype. Results indicate strong poisoning prevention guarantees at modest overhead applicable even to MCU-based resource-constrained IoT devices.

\section*{Acknowledgements}

Norrathep
Rattanavipanon was supported by
the National Science, Research and Innovation Fund (NSRF)
and Prince of Songkla University (Grant No. COC6701016S) and the NSRF via the Program Management Unit
for Human Resources \& Institutional Development, Research and Innovation [grant
number B13F670122].
Ivan De Oliveira Nunes was supported
by NSF Award SaTC- 2245531.

\bibliographystyle{plain}
\bibliography{references}

\begin{thebibliography}{10}

\bibitem{cflat}
Tigist Abera, N~Asokan, Lucas Davi, Jan-Erik Ekberg, Thomas Nyman, Andrew Paverd, Ahmad-Reza Sadeghi, and Gene Tsudik.
\newblock C-flat: control-flow attestation for embedded systems software.
\newblock In {\em ACM CCS}, 2016.

\bibitem{ammar2024sok}
Mahmoud Ammar, Adam Caulfield, and Ivan De~Oliveira Nunes.
\newblock Sok: Integrity, attestation, and auditing of program execution.
\newblock In {\em 2025 IEEE Symposium on Security and Privacy (SP)}, pages 77--77. IEEE Computer Society, 2024.

\bibitem{arbaugh1997secure}
William~A Arbaugh, David~J Farber, and Jonathan~M Smith.
\newblock A secure and reliable bootstrap architecture.
\newblock In {\em IEEE Symposium on Security and Privacy}, 1997.

\bibitem{blanchard2017machine}
Peva Blanchard, El~Mahdi El~Mhamdi, Rachid Guerraoui, and Julien Stainer.
\newblock Machine learning with adversaries: Byzantine tolerant gradient descent.
\newblock {\em Advances in neural information processing systems}, 30, 2017.

\bibitem{tytan}
Ferdinand Brasser, Brahim El~Mahjoub, Ahmad-Reza Sadeghi, Christian Wachsmann, and Patrick Koeberl.
\newblock Tytan: Tiny trust anchor for tiny devices.
\newblock In {\em DAC}, 2015.

\bibitem{cao2019understanding}
Di~Cao, Shan Chang, Zhijian Lin, Guohua Liu, and Donghong Sun.
\newblock Understanding distributed poisoning attack in federated learning.
\newblock In {\em 2019 IEEE 25th International Conference on Parallel and Distributed Systems (ICPADS)}, pages 233--239. IEEE, 2019.

\bibitem{cao2021data}
Xiaoyu Cao, Jinyuan Jia, and Neil~Zhenqiang Gong.
\newblock Data poisoning attacks to local differential privacy protocols.
\newblock In {\em 30th USENIX Security Symposium (USENIX Security 21)}, pages 947--964, 2021.

\bibitem{asap}
Adam Caulfield, Norrathep Rattanavipanon, and Ivan De~Oliveira~Nunes.
\newblock Asap: Reconciling asynchronous real-time operations and proofs of execution in simple embedded systems.
\newblock In {\em DAC}, 2022.

\bibitem{acfa}
Adam Caulfield, Norrathep Rattanavipanon, and Ivan De~Oliveira Nunes.
\newblock Acfa: Secure runtime auditing \& guaranteed device healing via active control flow attestation.
\newblock In {\em USENIX Security}, 2023.

\bibitem{celik2018soteria}
Z~Berkay Celik, Patrick McDaniel, and Gang Tan.
\newblock Soteria: Automated {IoT} safety and security analysis.
\newblock In {\em USENIX ATC 18}, 2018.

\bibitem{chen2017distributed}
Yudong Chen, Lili Su, and Jiaming Xu.
\newblock Distributed statistical machine learning in adversarial settings: Byzantine gradient descent.
\newblock {\em Proceedings of the ACM on Measurement and Analysis of Computing Systems}, 1(2):1--25, 2017.

\bibitem{vrased}
Ivan De~Oliveira~Nunes, Karim Eldefrawy, Norrathep Rattanavipanon, Michael Steiner, and Gene Tsudik.
\newblock {VRASED}: A verified hardware/software co-design for remote attestation.
\newblock In {\em 28th USENIX Security Symposium (USENIX Security 19)}, pages 1429--1446, 2019.

\bibitem{apex}
Ivan De~Oliveira~Nunes, Karim Eldefrawy, Norrathep Rattanavipanon, and Gene Tsudik.
\newblock {APEX}: A verified architecture for proofs of execution on remote devices under full software compromise.
\newblock In {\em 29th USENIX Security Symposium (USENIX Security 20)}, 2020.

\bibitem{tinycfa}
Ivan De~Oliveira~Nunes, Sashidhar Jakkamsetti, and Gene Tsudik.
\newblock Tiny-cfa: Minimalistic control-flow attestation using verified proofs of execution.
\newblock In {\em 2021 Design, Automation \& Test in Europe Conference \& Exhibition (DATE)}, pages 641--646. IEEE, 2021.

\bibitem{dialed}
Ivan De~Oliveira Nunes~et al.
\newblock Dialed: Data integrity attestation for low-end embedded devices.
\newblock In {\em 2021 58th ACM/IEEE Design Automation Conference (DAC)}, pages 313--318. IEEE, 2021.

\bibitem{SA4P}
Piet De~Vaere.
\newblock {\em Fine-Grained Access Control For Sensors, Actuators, and Automation Networks}.
\newblock PhD thesis, ETH Zurich, 2023.

\bibitem{de2024sa}
Piet De~Vaere, Felix St{\"o}ger, Adrian Perrig, and Gene Tsudik.
\newblock The sa4p framework: Sensing and actuation as a privilege.
\newblock {\em ACM AsiaCCS}, 2024.

\bibitem{dessouky2018litehax}
Ghada Dessouky, Tigist Abera, Ahmad Ibrahim, and Ahmad-Reza Sadeghi.
\newblock Litehax: lightweight hardware-assisted attestation of program execution.
\newblock In {\em 2018 IEEE/ACM International Conference on Computer-Aided Design (ICCAD)}, pages 1--8. IEEE, 2018.

\bibitem{lofat}
Ghada Dessouky, Shaza Zeitouni, Thomas Nyman, Andrew Paverd, Lucas Davi, Patrick Koeberl, N~Asokan, and Ahmad-Reza Sadeghi.
\newblock Lo-fat: Low-overhead control flow attestation in hardware.
\newblock In {\em Proceedings of the 54th Annual Design Automation Conference 2017}, pages 1--6, 2017.

\bibitem{ding2017collecting}
Bolin Ding~et al.
\newblock Collecting telemetry data privately.
\newblock {\em Advances in Neural Information Processing Systems}, 30, 2017.

\bibitem{hydra}
Karim Eldefrawy, Norrathep Rattanavipanon, and Gene Tsudik.
\newblock Hydra: hybrid design for remote attestation (using a formally verified microkernel).
\newblock In {\em ACM WiSec}, 2017.

\bibitem{smart}
Karim Eldefrawy, Gene Tsudik, Aur{\'e}lien Francillon, and Daniele Perito.
\newblock {SMART}: Secure and minimal architecture for (establishing dynamic) root of trust.
\newblock In {\em NDSS}, volume~12, pages 1--15, 2012.

\bibitem{erlingsson2014rappor}
{\'U}lfar Erlingsson, Vasyl Pihur, and Aleksandra Korolova.
\newblock Rappor: Randomized aggregatable privacy-preserving ordinal response.
\newblock In {\em Proceedings of the 2014 ACM SIGSAC conference on computer and communications security}, pages 1054--1067, 2014.

\bibitem{fang2020local}
Minghong Fang, Xiaoyu Cao, Jinyuan Jia, and Neil Gong.
\newblock Local model poisoning attacks to $\{$Byzantine-Robust$\}$ federated learning.
\newblock In {\em USENIX Security}, 2020.

\bibitem{feng2014robust}
Jiashi Feng, Huan Xu, Shie Mannor, and Shuicheng Yan.
\newblock Robust logistic regression and classification.
\newblock {\em Advances in neural information processing systems}, 27, 2014.

\bibitem{gu2019badnets}
Tianyu Gu, Kang Liu, Brendan Dolan-Gavitt, and Siddharth Garg.
\newblock Badnets: Evaluating backdooring attacks on deep neural networks.
\newblock {\em IEEE Access}, 7:47230--47244, 2019.

\bibitem{guerraoui2018hidden}
Rachid Guerraoui, S{\'e}bastien Rouault, et~al.
\newblock The hidden vulnerability of distributed learning in byzantium.
\newblock In {\em International Conference on Machine Learning}, pages 3521--3530. PMLR, 2018.

\bibitem{hashemi2021byzantine}
Hanieh Hashemi, Yongqin Wang, Chuan Guo, and Murali Annavaram.
\newblock Byzantine-robust and privacy-preserving framework for fedml.
\newblock {\em arXiv preprint arXiv:2105.02295}, 2021.

\bibitem{cambridgeanalytica}
Joanne Hinds, Emma~J Williams, and Adam~N Joinson.
\newblock “it wouldn't happen to me”: Privacy concerns and perspectives following the cambridge analytica scandal.
\newblock {\em International Journal of Human-Computer Studies}, 143:102498, 2020.

\bibitem{GDPR3}
Fran{\c{c}}ois Hublet, David Basin, and Srdjan Krsti{\'c}.
\newblock Enforcing the gdpr.
\newblock In {\em European Symposium on Research in Computer Security}, pages 400--422. Springer, 2023.

\bibitem{hulsing2016armed}
Andreas H{\"u}lsing, Joost Rijneveld, and Peter Schwabe.
\newblock Armed sphincs: Computing a 41 kb signature in 16 kb of ram.
\newblock In {\em IACR International Conference on Practice and Theory in Public-Key Cryptography}, 2016.

\bibitem{jagielski2018manipulating}
Matthew Jagielski, Alina Oprea, Battista Biggio, Chang Liu, Cristina Nita-Rotaru, and Bo~Li.
\newblock Manipulating machine learning: Poisoning attacks and countermeasures for regression learning.
\newblock In {\em 2018 IEEE symposium on security and privacy (SP)}, pages 19--35. IEEE, 2018.

\bibitem{viceroy}
Scott Jordan, Yoshimichi Nakatsuka, Ercan Ozturk, Andrew Paverd, and Gene Tsudik.
\newblock Viceroy: Gdpr-/ccpa-compliant enforcement of verifiable accountless consumer requests.
\newblock {\em NDSS}, 2023.

\bibitem{trustlite}
Patrick Koeberl, Steffen Schulz, Ahmad-Reza Sadeghi, and Vijay Varadharajan.
\newblock {TrustLite}: A security architecture for tiny embedded devices.
\newblock In {\em EuroSys}. ACM, 2014.

\bibitem{konevcny2016federated}
Jakub Kone{\v{c}}n{\`y}, H~Brendan McMahan, Felix~X Yu, Peter Richt{\'a}rik, Ananda~Theertha Suresh, and Dave Bacon.
\newblock Federated learning: Strategies for improving communication efficiency.
\newblock {\em arXiv preprint arXiv:1610.05492}, 2016.

\bibitem{GDPR1}
He~Li, Lu~Yu, and Wu~He.
\newblock The impact of gdpr on global technology development, 2019.

\bibitem{fedprox}
Tian Li, Anit~Kumar Sahu, Manzil Zaheer, Maziar Sanjabi, Ameet Talwalkar, and Virginia Smith.
\newblock Federated optimization in heterogeneous networks.
\newblock {\em Proceedings of Machine learning and systems}, 2:429--450, 2020.

\bibitem{li2023fine}
Xiaoguang Li, Ninghui Li, Wenhai Sun, Neil~Zhenqiang Gong, and Hui Li.
\newblock Fine-grained poisoning attack to local differential privacy protocols for mean and variance estimation.
\newblock In {\em USENIX Security}, 2023.

\bibitem{ma2022shieldfl}
Zhuoran Ma, Jianfeng Ma, Yinbin Miao, Yingjiu Li, and Robert~H Deng.
\newblock Shieldfl: Mitigating model poisoning attacks in privacy-preserving federated learning.
\newblock {\em IEEE Transactions on Information Forensics and Security}, 17:1639--1654, 2022.

\bibitem{flicker}
Jonathan~M McCune, Bryan~J Parno, Adrian Perrig, Michael~K Reiter, and Hiroshi Isozaki.
\newblock Flicker: An execution infrastructure for tcb minimization.
\newblock In {\em ACM EuroSys}, 2008.

\bibitem{mcdaniel2009security}
Patrick McDaniel and Stephen McLaughlin.
\newblock Security and privacy challenges in the smart grid.
\newblock {\em IEEE security \& privacy}, 7(3):75--77, 2009.

\bibitem{messaoud2022shielding}
Aghiles~Ait Messaoud, Sonia~Ben Mokhtar, Vlad Nitu, and Valerio Schiavoni.
\newblock Shielding federated learning systems against inference attacks with arm trustzone.
\newblock In {\em Proceedings of the 23rd ACM/IFIP International Middleware Conference}, pages 335--348, 2022.

\bibitem{mo2021ppfl}
Fan Mo, Hamed Haddadi, Kleomenis Katevas, Eduard Marin, Diego Perino, and Nicolas Kourtellis.
\newblock Ppfl: Privacy-preserving federated learning with trusted execution environments.
\newblock In {\em Proceedings of the 19th annual international conference on mobile systems, applications, and services}, pages 94--108, 2021.

\bibitem{moharana2017secure}
Soumya~Ranjan Moharana, Vijay~Kumar Jha, Anurag Satpathy, Sourav~Kanti Addya, Ashok~Kumar Turuk, and Banshidhar Majhi.
\newblock Secure key-distribution in iot cloud networks.
\newblock In {\em 2017 Third International Conference on Sensing, Signal Processing and Security (ICSSS)}, pages 197--202. IEEE, 2017.

\bibitem{molina2010private}
Andr{\'e}s Molina-Markham, Prashant Shenoy, Kevin Fu, Emmanuel Cecchet, and David Irwin.
\newblock Private memoirs of a smart meter.
\newblock In {\em Proceedings of the 2nd ACM workshop on embedded sensing systems for energy-efficiency in building}, pages 61--66, 2010.

\bibitem{mossinger2016towards}
Max M{\"o}ssinger, Benedikt Petschkuhn, Johannes Bauer, Ralf~C Staudemeyer, Marcin W{\'o}jcik, and Henrich~C P{\"o}hls.
\newblock Towards quantifying the cost of a secure iot: Overhead and energy consumption of ecc signatures on an arm-based device.
\newblock In {\em 2016 IEEE 17th International Symposium on A World of Wireless, Mobile and Multimedia Networks (WoWMoM)}, pages 1--6. IEEE, 2016.

\bibitem{cve_2017_14201}
{National Vulnerability Database}.
\newblock {CVE-2017-14201}, 2017.
\newblock Accessed: 2024-10-10.

\bibitem{cve_2020_10023}
{National Vulnerability Database}.
\newblock {CVE-2020-10023}, 2020.
\newblock Accessed: 2024-10-10.

\bibitem{iscflat}
Antonio~Joia Neto and Ivan De~Oliveira Nunes.
\newblock Isc-flat: On the conflict between control flow attestation and real-time operations.
\newblock In {\em RTAS}, 2023.

\bibitem{niederhagen2022streaming}
Ruben Niederhagen, Johannes Roth, and Julian W{\"a}lde.
\newblock Streaming sphincs+ for embedded devices using the example of tpms.
\newblock In {\em International Conference on Cryptology in Africa}, 2022.

\bibitem{Sancus17}
Job Noorman, Jo~Van Bulck, Jan~Tobias M{\"u}hlberg, Frank Piessens, Pieter Maene, Bart Preneel, Ingrid Verbauwhede, Johannes G{\"o}tzfried, Tilo M{\"u}ller, and Felix Freiling.
\newblock Sancus 2.0: A low-cost security architecture for iot devices.
\newblock {\em ACM TOPS}, 20(3):1--33, 2017.

\bibitem{versa}
Ivan De~Oliveira Nunes, Seoyeon Hwang, Sashidhar Jakkamsetti, and Gene Tsudik.
\newblock Privacy-from-birth: Protecting sensed data from malicious sensors with versa.
\newblock In {\em IEEE Symposium on Security and Privacy}, 2022.

\bibitem{mini-survey}
Ivan De~Oliveira Nunes, Sashidhar Jakkamsetti, Norrathep Rattanavipanon, and Gene Tsudik.
\newblock Towards remotely verifiable software integrity in resource-constrained iot devices.
\newblock {\em IEEE Communications Magazine}, 2024.

\bibitem{rares}
Avani Dave Nilanjan Banerjee~Chintan Patel.
\newblock Rares: Runtime attack resilient embedded system design using verified proof-of-execution.
\newblock {\em arXiv preprint arXiv:2305.03266}, 2023.

\bibitem{trustzone}
Sandro Pinto and Nuno Santos.
\newblock Demystifying arm trustzone: A comprehensive survey.
\newblock {\em ACM computing surveys (CSUR)}, 51(6):1--36, 2019.

\bibitem{queyrut2023mitigating}
Simon Queyrut, Valerio Schiavoni, and Pascal Felber.
\newblock Mitigating adversarial attacks in federated learning with trusted execution environments.
\newblock In {\em 2023 IEEE 43rd International Conference on Distributed Computing Systems (ICDCS)}, pages 626--637. IEEE, 2023.

\bibitem{repo}
Norrathep Rattanavipanon and Ivan De~Oliveira Nunes.
\newblock Slapp repo.
\newblock \url{https://github.com/norrathep/SLAPP}.

\bibitem{ravi2004tamper}
Srivaths Ravi, Anand Raghunathan, and Srimat Chakradhar.
\newblock Tamper resistance mechanisms for secure embedded systems.
\newblock In {\em VLSI Design}, 2004.

\bibitem{rieger2022crowdguard}
Phillip Rieger, Torsten Krau{\ss}, Markus Miettinen, Alexandra Dmitrienko, and Ahmad-Reza Sadeghi.
\newblock Crowdguard: Federated backdoor detection in federated learning.
\newblock {\em arXiv preprint arXiv:2210.07714}, 2022.

\bibitem{silde2019comparative}
Tjerand Silde.
\newblock Comparative study of ecc libraries for embedded devices.
\newblock {\em Norwegian University of Science and Technology, Tech. Rep}, 2019.

\bibitem{nucleo}
{STMicroelectronics}.
\newblock Nucleo-l552ze-q.
\newblock \url{https://estore.st.com/en/nucleo-l552ze-q-cpn.html}.

\bibitem{sun2021fl}
Jingwei Sun, Ang Li, Louis DiValentin, Amin Hassanzadeh, Yiran Chen, and Hai Li.
\newblock Fl-wbc: Enhancing robustness against model poisoning attacks in federated learning from a client perspective.
\newblock {\em Advances in Neural Information Processing Systems}, 34:12613--12624, 2021.

\bibitem{oat}
Zhichuang Sun~et al.
\newblock Oat: Attesting operation integrity of embedded devices.
\newblock In {\em IEEE S\&P}, 2020.

\bibitem{taik2020electrical}
Afaf Ta{\"\i}k and Soumaya Cherkaoui.
\newblock Electrical load forecasting using edge computing and federated learning.
\newblock In {\em IEEE international conference on communications (ICC)}, pages 1--6. IEEE, 2020.

\bibitem{scarr}
Flavio Toffalini, Eleonora Losiouk, Andrea Biondo, Jianying Zhou, and Mauro Conti.
\newblock {ScaRR}: Scalable runtime remote attestation for complex systems.
\newblock In {\em 22nd International Symposium on Research in Attacks, Intrusions and Defenses (RAID 2019)}, pages 121--134, 2019.

\bibitem{tolpegin2020data}
Vale Tolpegin, Stacey Truex, Mehmet~Emre Gursoy, and Ling Liu.
\newblock Data poisoning attacks against federated learning systems.
\newblock In {\em ESORICS}, 2020.

\bibitem{tramer2020adaptive}
Florian Tramer, Nicholas Carlini, Wieland Brendel, and Aleksander Madry.
\newblock On adaptive attacks to adversarial example defenses.
\newblock {\em Advances in neural information processing systems}, 33:1633--1645, 2020.

\bibitem{tpm}
{Trusted Computing Group.}
\newblock Trusted platform module (tpm), 2017.

\bibitem{tsudik2024staving}
Gene Tsudik.
\newblock Staving off the iot armageddon.
\newblock In {\em Proceedings of the 2024 on ACM SIGSAC Conference on Computer and Communications Security}, pages 2--3, 2024.

\bibitem{wang2023ari}
Jinwen Wang, Yujie Wang, Ao~Li, Yang Xiao, Ruide Zhang, Wenjing Lou, Y~Thomas Hou, and Ning Zhang.
\newblock {ARI}: Attestation of real-time mission execution integrity.
\newblock In {\em 32nd USENIX Security Symposium (USENIX Security 23)}, pages 2761--2778, 2023.

\bibitem{wenger2021backdoor}
Emily Wenger, Josephine Passananti, Arjun~Nitin Bhagoji, Yuanshun Yao, Haitao Zheng, and Ben~Y Zhao.
\newblock Backdoor attacks against deep learning systems in the physical world.
\newblock In {\em Proceedings of the IEEE/CVF conference on computer vision and pattern recognition}, pages 6206--6215, 2021.

\bibitem{ebfa}
Jingyi Yao, Chen Song, Hongjia Li, Yuxiang Wang, Qian Yang, and Liming Wang.
\newblock An enclave-aided byzantine-robust federated aggregation framework.
\newblock In {\em 2024 IEEE Wireless Communications and Networking Conference (WCNC)}, pages 1--6. IEEE, 2024.

\bibitem{GDPR2}
Razieh~Nokhbeh Zaeem and K~Suzanne Barber.
\newblock The effect of the gdpr on privacy policies: Recent progress and future promise.
\newblock {\em ACM Transactions on Management Information Systems (TMIS)}, 12(1):1--20, 2020.

\bibitem{zeitouni2017atrium}
Shaza Zeitouni, Ghada Dessouky, Orlando Arias, Dean Sullivan, Ahmad Ibrahim, Yier Jin, and Ahmad-Reza Sadeghi.
\newblock Atrium: Runtime attestation resilient under memory attacks.
\newblock In {\em ICCAD}, 2017.

\bibitem{recfa}
Yumei Zhang, Xinzhi Liu, Cong Sun, Dongrui Zeng, Gang Tan, Xiao Kan, and Siqi Ma.
\newblock {ReCFA}: resilient control-flow attestation.
\newblock In {\em Annual Computer Security Applications Conference}, pages 311--322, 2021.

\bibitem{zhao2018federated}
Yue Zhao, Meng Li, Liangzhen Lai, Naveen Suda, Damon Civin, and Vikas Chandra.
\newblock Federated learning with non-iid data.
\newblock {\em arXiv preprint arXiv:1806.00582}, 2018.

\end{thebibliography}

\revision{
\begin{IEEEbiography}
[{\includegraphics[width=1in,height=1.25in,clip,keepaspectratio]{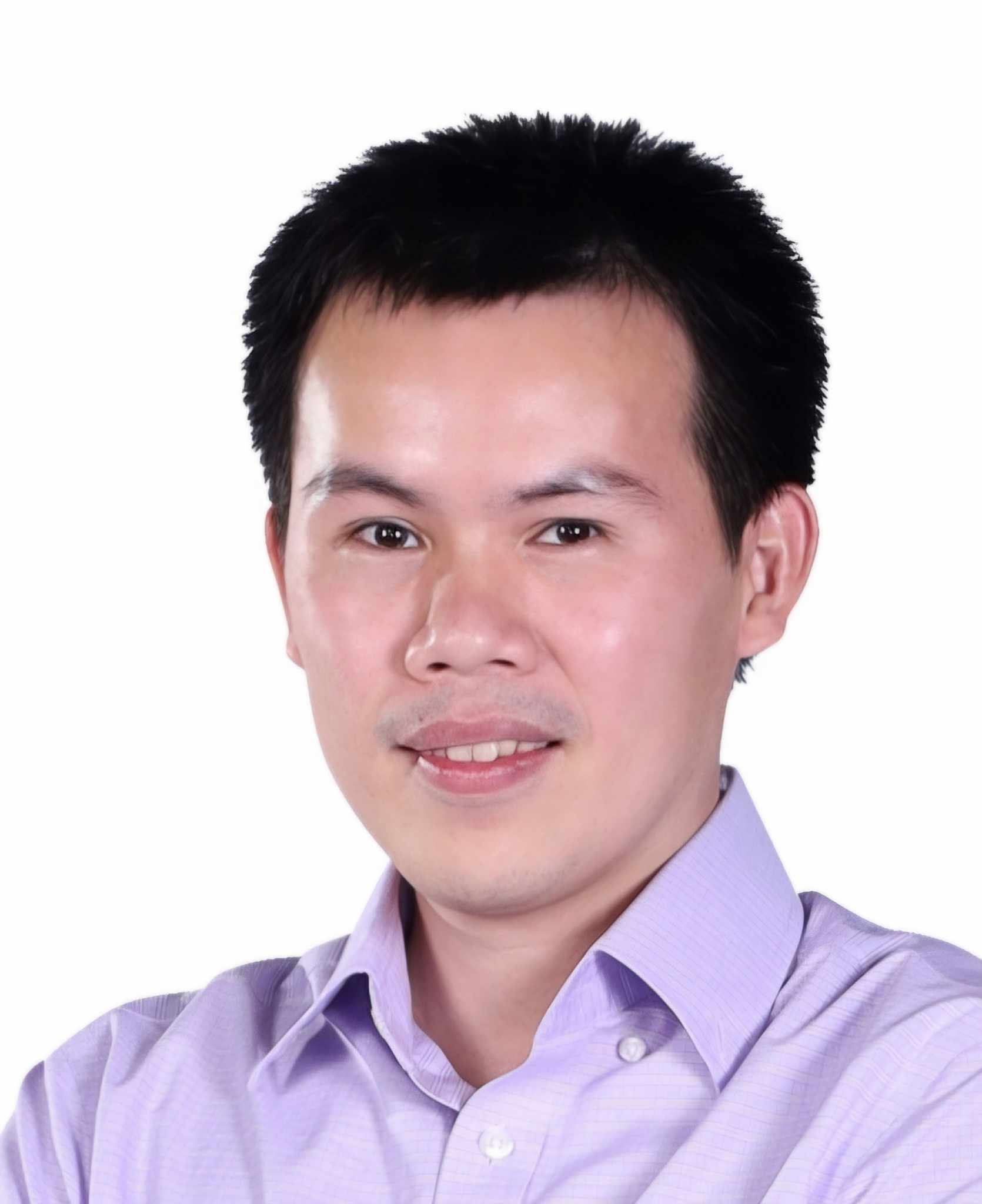}}]{Norrathep Rattanavipanon} 
received his Ph.D. in Computer Science from the University of California, Irvine in 2019. Currently, he is an Assistant Professor with the College of Computing, Prince of Songkla University, Phuket Campus. 
His research interests lie in the area of security and privacy, particularly in embedded systems and IoT security, software and binary analysis, and security/privacy in machine learning systems.
\end{IEEEbiography}
\vspace{-5 mm}
\begin{IEEEbiography}[{\includegraphics[width=1in,height=1.25in,clip,keepaspectratio]{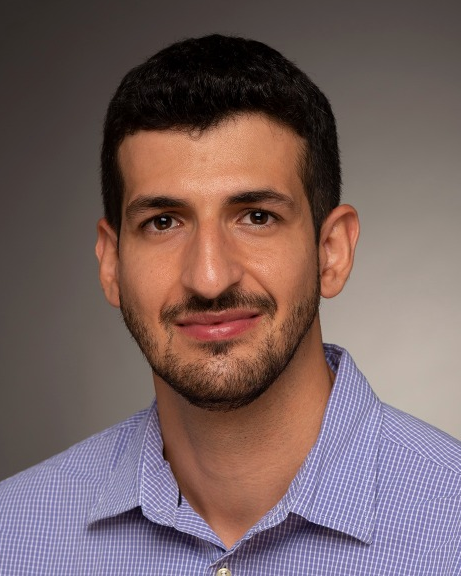}}]{Ivan De Oliveira Nunes}

Ivan is an Assistant Professor at the University of Zurich (UZH). Prior to joining UZH, he was an Assistant Professor at the Rochester Institute of Technology. He earned his Ph.D. from the University of California, Irvine. His research interests include Security \& Privacy, Computer Networking, Computing Systems, and particularly the intersections of these fields.
\end{IEEEbiography}
}

\end{document}